 \documentclass[sigconf]{acmart} 
\AtBeginDocument{%
  \providecommand\BibTeX{{%
    \normalfont B\kern-0.5em{\scshape i\kern-0.25em b}\kern-0.8em\TeX}}}

\copyrightyear{2024}
\acmYear{2024}
\setcopyright{rightsretained}
\acmConference[CHI '24]{Proceedings of the CHI Conference on Human Factors in Computing Systems}{May 11--16, 2024}{Honolulu, HI, USA}
\acmBooktitle{Proceedings of the CHI Conference on Human Factors in Computing Systems (CHI '24), May 11--16, 2024, Honolulu, HI, USA}
\acmDOI{10.1145/3613904.3642450}
\acmISBN{979-8-4007-0330-0/24/05}

\usepackage{subcaption}
\usepackage{xcolor}

\begin{document}

%%
%% The "title" command has an optional parameter,
%% allowing the author to define a "short title" to be used in page headers.
\title[Memoro: Using LLMs to Realize a Concise Interface for Real-Time Memory Augmentation]{Memoro: Using Large Language Models to Realize a Concise Interface for Real-Time Memory Augmentation}

%{Echoic: Wearable Interface for Seamless Memory Retrieval}

%% The "author" command and its associated commands are used to define
%% the authors and their affiliations.
%% Of note is the shared affiliation of the first two authors, and the
%% "authornote" and "authornotemark" commands
%% used to denote shared contribution to the research.
%%
%% The "author" command and its associated commands are used to define
%% the authors and their affiliations.
%% Of note is the shared affiliation of the first two authors, and the
%% "authornote" and "authornotemark" commands
%% used to denote shared contribution to the research.

\author{Wazeer Zulfikar}
\affiliation{%
  \institution{MIT Media Lab}
  \city{Cambridge}
  \country{USA}}
\email{wazeer@media.mit.edu}

\author{Samantha Chan}
\affiliation{%
  \institution{MIT Media Lab}
  \city{Cambridge}
  \country{USA}}
  \email{swtchan@media.mit.edu}

\author{Pattie Maes}
\affiliation{%
  \institution{MIT Media Lab}
  \city{Cambridge}
\country{USA}}
  \email{pattie@media.mit.edu}

%%
%% By default, the full list of authors will be used in the page
%% headers. Often, this list is too long, and will overlap
%% other information printed in the page headers. This command allows
%% the author to define a more concise list
%% of authors' names for this purpose.
\renewcommand{\shortauthors}{Zulfikar, et al.}

%%
%% The abstract is a short summary of the work to be presented in the
%% article.
\begin{abstract}
  % A clear and well-documented \LaTeX\ document is presented as an
  % article formatted for publication by ACM in a conference proceedings
  % or journal publication. Based on the ``acmart'' document class, this
  % article presents and explains many of the common variations, as well
  % as many of the formatting elements an author may use in the
  % preparation of the documentation of their work.

People have to remember an ever-expanding volume of information. Wearables that use information capture and retrieval for memory augmentation can help but can be disruptive and cumbersome in real-world tasks, such as in social settings. To address this, we developed Memoro, a wearable audio-based memory assistant with a concise user interface. Memoro uses a large language model (LLM) to infer the user’s memory needs in a conversational context, semantically search memories, and present minimal suggestions. The assistant has two interaction modes: Query Mode for voicing queries and Queryless Mode for on-demand predictive assistance, without explicit query. Our study of (N=20) participants engaged in a real-time conversation, demonstrated that using Memoro reduced device interaction time and increased recall confidence while preserving conversational quality. We report quantitative results and discuss the preferences and experiences of users. This work contributes towards utilizing LLMs to design wearable memory augmentation systems that are minimally disruptive.
\end{abstract}


%%
%% Keywords. The author(s) should pick words that accurately describe
%% the work being presented. Separate the keywords with commas.

%% A "teaser" image appears between the author and affiliation
%% information and the body of the document, and typically spans the
%% page.
\begin{teaserfigure}
  \includegraphics[width=\linewidth]{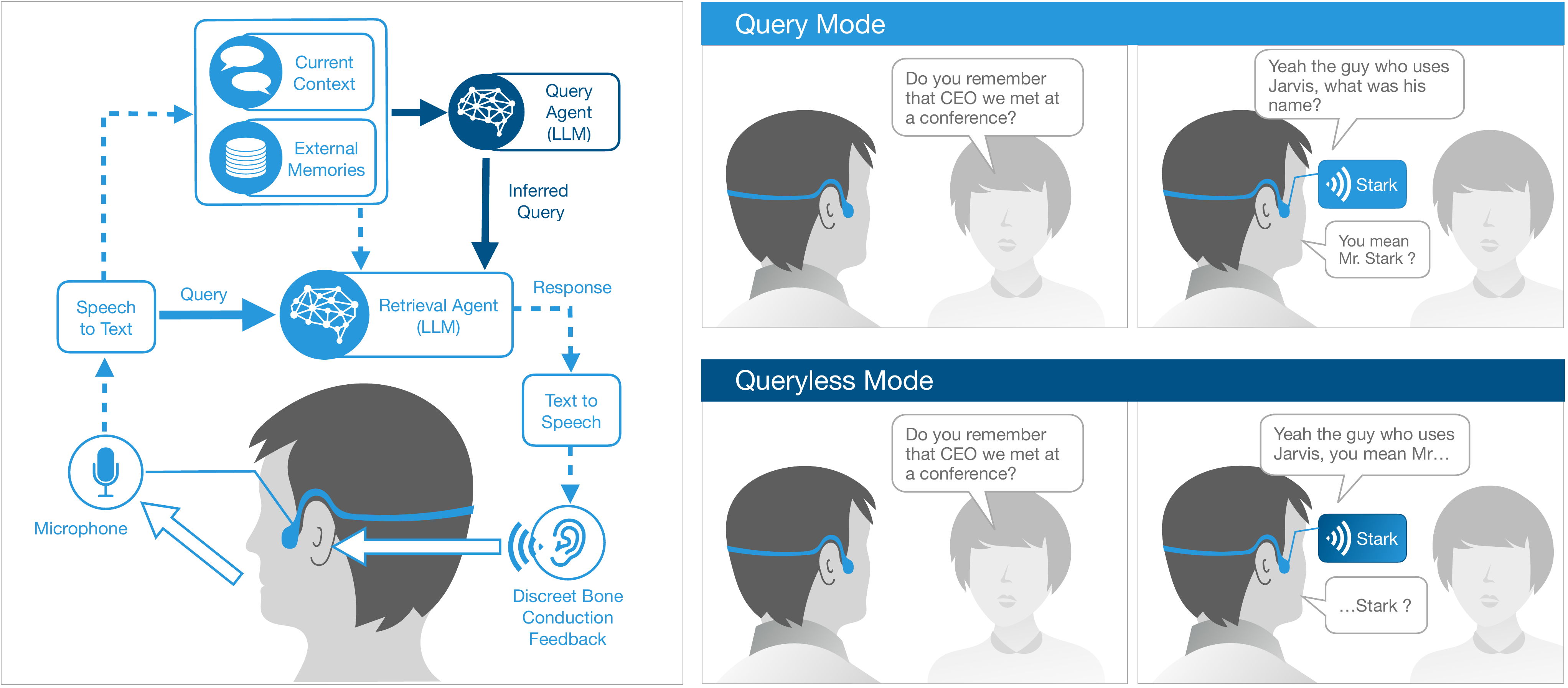}
  \caption{Architecture of Memoro and its two interaction modes. (Left) System architecture of the memory assistant. (Right) Two interaction modes: (1) \textit{Query Mode} where the user can ask contextual questions (2) \textit{Queryless Mode} where the user can request predictive assistance and skip query formation. In both modes, responses are discreetly played back to the user using a bone conduction headset.}
  \Description{Teaser}
  \label{fig:teaser}
\end{teaserfigure}

\received{15 May 2023}
% \received[revised]{12 March 2009}
% \received[accepted]{5 June 2009}

%%
%% This command processes the author and affiliation and title
%% information and builds the first part of the formatted document.
\maketitle

\section{Introduction}

Memory plays an essential role in people's lives, whether in communication, learning, decision-making, or maintaining relationships \cite{baddeley2015memory, lynch1991memory}. However, memory is imperfect and error-prone due to factors such as lack of sleep, stress, and divided attention \cite{rasch2013sleep, schacter1999seven}. Furthermore, neurological disorders related to memory loss, such as dementia, are rising as populations in many parts of the world grow older \cite{PRINCE201363}. 

Memory augmentation and information retrieval systems have been of key interest to the HCI community over the past several decades as tools to address these growing challenges. Since Vannevar Bush's conception of the Memex in 1945 \cite{bush1945we}, there has been extensive work on systems and devices to extend our memory \cite{Rhodes2000, Chan2020, chan2019prospero, ksibi2021overview} such as lifelogging systems that continuously record the user's media and signals \cite{Mann1996, Vemuri2004, jiang2019memento}, and just-in-time information retrieval systems \cite{khan2019, Bermejo2020, Rhodes2000, marmasse1999commotion, Devaul2004} that provide relevant information based on the user's context. While these wearable systems demonstrate the capabilities of users to retrieve vast amounts of information, limited research exists on designing interfaces that enable the retrieval of information in a minimally disruptive way when the user is already engaged in a primary task, which is often the case with wearables. 

We define minimal disruption for a memory augmentation interface as (1) requiring minimal input from the user to request information, i.e., the input the user gives is short, and (2) providing minimal output, namely the suggestion or response provided by the augmentation system is the smallest amount of information that will give the user the information they need. The minimal disruption design consideration is critical for the usability of wearable memory augmentation systems \cite{gelonch2019acceptability}, especially in social settings that are attention-demanding and where incidentally the highest number of memory lapses occur \cite{Lynn2013}, such as conversations.

Therefore, an important challenge for the design of wearable memory augmentation systems is that of a seamless, user-friendly, and concise search interface \cite{gelonch2019acceptability} to keep disruption to the user's primary task minimal. Incorporating context awareness can reduce or, as we show in this paper, even completely eliminate the query input, allowing users to skip posing an explicit, comprehensive retrieval query, as the system can directly infer the user's specific memory needs. Recent developments in large language models (LLMs) have improved capabilities in understanding conversational context in natural settings \cite{brown2020language, wei2021finetuned} and enable more flexible search queries using alternative phrases \cite{lewis2021retrievalaugmented}. They also enable the shortening of answers \cite{goyal2022news} for succinct suggestions. This highlights the opportunity to leverage LLMs to design easy-to-use and minimally disruptive interfaces.

In this paper, we aim to answer the following research questions
%\\
\begin{itemize}

    \item \textbf{RQ1.} How can we design a seamless wearable memory assistant using LLMs to reduce disruption to the primary task with minimal and effective input and output? 
    \item \textbf{RQ2.} What are the effects of using the memory augmentation system during the primary task of a real-time conversation across metrics such as quality of conversation, performance, and task load?
    \item \textbf{RQ3.} How do context awareness and conciseness affect the system's usability, user perception and experience?

\end{itemize} 

We developed a minimally disruptive audio-based wearable assistant, Memoro, that uses LLMs to aid the user in retrieving relevant information from previously recorded personal data through concise suggestions. Memoro continuously transcribes and encodes audio data from conversations the user engages in. The memory assistant has two modes of interaction for retrieval: Query Mode, where the user voices their natural language query, and Queryless Mode where the user is presented with a suggestion relevant to the current conversational context without having to explicitly query the system. Both modes provide minimal memory responses to the user (see Figure \ref{fig:teaser}). In terms of hardware form factor, Memoro uses a light-weight, bone-conduction headset for unobstructed and private responses.

% In our within-subjects study with 20 participants, users engaged in a social interaction task under four conditions:
% \begin{itemize}
% \item  ``No System'' which was the control condition  
% \item ``Baseline'' LLM system with explicit query and raw, full-length answers
% \item  \textit{Query Mode} of Memoro with explicit contextual query and concise answers
% \item  \textit{Queryless Mode} of Memoro with no query and concise answers. 
% \end{itemize}
% The No System condition was used to compare the measures of conversational quality and task load of using Memoro, while the Baseline condition was used to measure the importance of contextual awareness and conciseness in the system. In addition, technical evaluations were conducted to measure the response accuracy and verbosity. The interaction modes were separately analyzed for a detailed evaluation.

To study the use of Memoro and its two query modes in the context of a real-time conversation, we conducted a study with N=20 participants. We found that the use of Memoro increased their recall confidence while preserving conversational quality. We also conducted a technical evaluation to measure the conciseness of input and output and the accuracy of the system responses. Most participants (15 of 20) expressed a preference for Memoro over no system and baseline (system without context awareness and conciseness), with 10 participants favoring the Queryless Mode. Participants elaborated upon their preferences and reservations, allowing for future design considerations. The highest-rated condition, Query Mode, achieved a mean usability score of 80.0, which falls between the good and excellent range \cite{bangor2009determining} and was significantly improved due to contextual awareness and conciseness as compared to the baseline. The goal of this paper is not to present a full-fledged memory augmentation system, but rather to evaluate whether LLMs can be used to make memory augmentation systems that are less disruptive.

In summary, the contributions of this paper are threefold:
\begin{enumerate}
    \item Design of a wearable memory assistant, called Memoro, focusing on minimal disruption to the user's primary real-world task by using conversational context and conciseness.
    \item Exploration of a query-less approach to eliminate query time and thereby increase seamless memory assistance by inferring the user's memory need.
    \item A within-subject user study showing that the proposed system has good usability and low interruption in a social task while preserving conversation quality and decreasing task load as compared to no system.
\end{enumerate}

\section{Related Work}
Our work is related to, and inspired by past work on wearable memory augmentation systems, context-aware agents in conversations, and large language models in virtual assistants.

\subsection{Wearable Memory Augmentation Systems}
Wearable memory augmentation has been a well-researched area since the 1990s when Mik Lamming coined the term "memory prosthesis" \cite{Lamming1994}. Since then, there have been various forms of memory augmentation systems, including reminder systems and lifelogging systems \cite{Lamming1994, Rhodes1997, Vemuri2004, Mann1996, hayes2004, harvey2016remembering, jiang2019memento, chan2020biosignal}. Lifelogging devices continuously capture signals such as audio, video, and biosignals resulting in a vast store of data. In the audio domain, Vemuri et al. \cite{Vemuri2004} introduced a personal audio memory aid that can record information and allow the user to search it using keywords. Hayes et al. \cite{hayes2004} showed the personal audio loop (PAL) as a ubiquitous service to recover audio content. Yamano and Itou \cite{yamano2009} and Shah et al. \cite{shah2012} recorded audio lifelogs using wearable microphones and experimented with different ways of browsing these lifelogs through a smartphone application. However, such types of browsing and keyword querying of audio data require a screen and, hence, use the users' visual focus and time to read the information provided. Gelonch et al found that an important factor in the acceptance of wearables in memory augmentation was the ease of use \cite{gelonch2019acceptability}. Furthermore, they were not designed to have quick and seamless interactions where disruption time during usage is critical, such as in conversations or driving.%* compared to visual wearable assistants (smart glasses etc.), we argue that those systems still need time to and people's visual focus to read the information provided by the systems.

Enabling voice-based interfaces for the users helps in memory retrieval from their lifelogs~\cite{pyae2018investigating}. Furthermore, voice interfaces can enable users to maintain high face focus and eye contact during conversations~\cite{cai2023paraglassmenu}. Therefore, we present a voice-based retrieval approach for an audio-based wearable memory assistant that can handle natural language queries with a focus on minimizing disruption to the primary task of the user. With concise responses from the assistant serving as memory suggestions, we aim to reduce device interaction time and preserve the quality of the primary task while using the system. Additionally, when the user is trying to retrieve specific details from a lifelog, we explore a method to allow users to skip having to form an explicit query by having the assistant infer their memory retrieval query based on the current context, as explained in the section below.

\subsection{Context-aware Agents in Conversations}
Just-in-time retrieval systems \cite{khan2019, Rhodes2000} aim to speed up the retrieval process by proactively retrieving relevant information from the database based on the user's current context. Social interactions such as conversations is a setting in which a majority of subjective memory complaints occur \cite{Lynn2013}. More recently, there has been growing work on real-time information access during conversations to bring filtered information to the user's attention to improve the quality of conversation \cite{Muller2017, kirilyuk2023visual, danry2020wearable, Andolina2018, Andolina2018Searchbot, Meurisch2020}.  Meurisch et al. \cite{Meurisch2020} conducted an in-the-wild study of systems with different proactivity levels. Muller et al. \cite{Muller2017} presented guidelines for the design of user interfaces for conversation support such as to provide means for fluid transition and re-engagement to ease the switch between information retrieval and the conversation. This informed the design of the interaction of Memoro to reduce query time and response duration. While there are several ways of providing proactive support, Liu et al. \cite{kirilyuk2023visual} show that a majority of users in a conversation preferred an on-demand suggestion interface over a fully proactive interface as it can be less distracting to the user experience. Wearable systems should minimize experiential disruptions to reduce users' explicit awareness of the system as this decreases cognitive load, and increases the sense of agency and sense of body-ownership~\cite{morris5wearable}. As minimizing distraction is central to the design of Memoro, this inspired our approach to providing on-demand predictive assistance, through the Queryless Mode, in memory retrieval. Understanding user intentions and conversational context is facilitated through recent advances in LLMs.

\subsection{Large Language Models in Virtual Assistants}
Virtual assistants are becoming increasingly important \cite{juniper2019} for information retrieval tasks that assist users. Guy \cite{guy2016} showed that the language of voice queries is closer to natural language than typed queries. Recent advances in natural language processing, particularly the development of LLMs, showed improved performance in question-answering tasks \cite{izacard2022atlas, rae2022scaling, brown2020language}. With the integration of language models in voice assistants, users can interact with systems using natural language. They can provide flexibility in user queries for different language use, such as synonyms, and alternative phrasings, and can compensate for inaccurate voice transcription due to the prerecorded priors \cite{vangysel2023}. This capacity is attributed to LLMs' ability to comprehend intentions and generate natural language in a contextualized manner. Further, vectorized embeddings of text generated by these models facilitate semantic search which enables diverse queries\cite{lewis2021retrievalaugmented}. For instance, while the recorded memory can be \textit{"He likes to hike and jog"}, a successful natural language voice query can be \textit{"What are his outdoor hobbies?"}, which has zero keyword matches. Furthermore, LLMs are adept at summarization tasks \cite{goyal2022news} aiding in providing minimal output to users in the concise interface. These concepts have not been explored in the context of wearable memory augmentation systems for improving usability during conversations. Hence, we leverage the capabilities of LLMs to power flexible search through memories and to interact with a voice-based assistant.

\section{System Design and Implementation}
Memoro, or "I remember" in Latin, is an audio-based memory assistant with a concise user interface. It continuously listens to the surrounding audio and encodes the raw speech transcriptions in memory, tagged by the timestamp at which it was transcribed {\color{black} and stored locally in the device, similar to  previous works}~\cite{Vemuri2004, hayes2004}. Whenever the user is in a primary task and has a real-time need for retrieval of information, they can trigger the system by pressing a ring button. The button informs the system that the user has a memory need. The button push can trigger one of two interaction modes:

\begin{enumerate}
    \item \textbf{Query Mode}: The user can explicitly query their Memoro system using natural language speech. If the user is in an ongoing conversation, the user can ask a brief question related to the conversation as the system is continuously listening, thus giving it conversational contextual awareness. For example, if the user is talking to a supermarket attendant and has said \textit{"I have bought eggs and bread"} in the conversation and wishes to remember the third thing they intended to purchase, they can hold the trigger button for Query Mode while asking \textit{"What was the third thing?"}. The system would then retrieve the answer, \textit{"Bananas"}, from the previously recorded memories. The retrieved answer is converted to audio using text-to-speech and played to the user through a bone-conduction headset.
    \item \textbf{Queryless Mode}: The user can also request predictive assistance, such that the system will infer the information that the user needs based on the current context and deliver the response without any explicit query from the user, similar to an autocomplete functionality. With the same example as above, after saying \textit{"I have bought eggs and bread but need to buy .."}, the user could trigger the Queryless Mode for the system by pressing the button which will based on understanding of the conversational context, infer the query, and respond with the suggestion \textit{"Bananas"} for the user to integrate into their incomplete sentence. 
\end{enumerate}

\begin{figure*}
  \includegraphics[width=\linewidth]{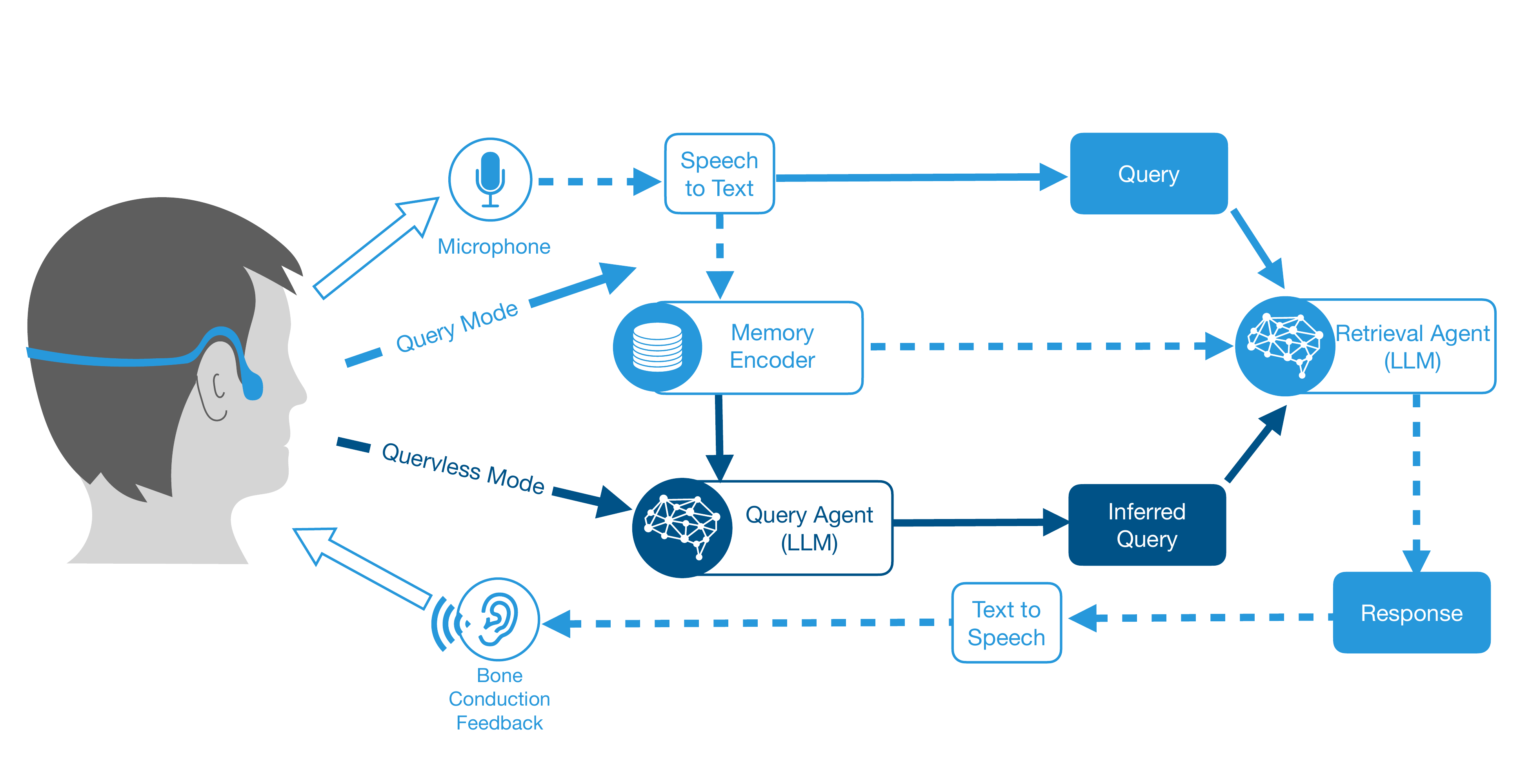}
  \caption{The closed loop system architecture has a memory encoder that is continuously updated using text-to-speech. The system can be configured to use query or queryless mode. In the Query Mode, the explicit query is voiced by the user, while in the Queryless Mode, the query agent infers the query. The query and memories are inputted to the retrieval agent, which returns a concise memory suggestion that is delivered to the user through bone conduction}
  \Description{Workflow of system}
  \label{fig:workflow}
\end{figure*}

Memoro has three components: the memory encoder, the retrieval agent, and the query agent. Figure~\ref{fig:workflow} shows an overview of the complete system architecture. 

The wearable platform consists of a commercial bone conduction headset that communicates with a smartphone or laptop. The bone conduction headset gives the user a parallel channel of audio \cite{moray1959attention, hunn2014hearables}, allowing them to have conversations with people while being able to hear audio responses from Memoro without impeding their field of view. The headset has an in-built microphone. Speech recognition is implemented using Google's Speech-to-Text API and speech synthesis of the text response from the memory assistant uses the Google Text-to-Speech API. The large language model used is OpenAI GPT3 (davinci-003) \cite{brown2020language} with a temperature of 0.

\subsection{Memory Encoder}
Auditory memories are stored using a two-step process. A continuous transcription is run on what the microphone picks up, including both the speech of the user and the conversation partner, under the assumption that privacy consent has been addressed. The transcription is first stored as the \textbf{Current Context} of the conversation. The current context is maintained in a fixed-sized buffer of the last $\alpha$ characters of data. We set $\alpha$ to 75 characters for capturing the most recent couple of sentences in the prototype but can be set larger to capture more context. The buffer is continuously updated by adding new information and removing information that is beyond the $\alpha$ threshold specified earlier. The set of information removed from the current context is chunked together into a single block and then encoded into the \textbf{External Memories} as a memory. 

Encoding of the memory is done using sentence embedding vectors of the text transcription of the full block. The embeddings capture the meaning of the memory enabling semantic search beyond keyword matching. Embeddings are calculated using pre-trained all-MiniLM-L6-v2 sentence transformer model \cite{reimers2019sentence} which maps sentences and paragraphs to a 384-dimensional dense vector space. Through these embeddings, the most semantically relevant memories containing the answer to the user query can be selected during retrieval. The embeddings, the text transcription, and the start timestamp for each memory block are stored using a vector database for faster retrieval \cite{kushilevitz1998efficient}. Figure \ref{fig:detailed_workflow} shows the encoding process of transcriptions into external memories.

\subsection{Retrieval Agent}
The aim of the retrieval agent is to take a query and respond with a concise answer from the user's encoded external memories, enabling the \textit{Query Mode}. It uses a method called retrieval augmented generation developed by Lewis et al\cite{lewis2021retrievalaugmented} and used in state of the art question-answering systems \cite{nassiri2023transformer, siriwardhana2023improving}.

% retrieval augmented generation
% Using vectorDB and sentence embeddings, prompt engineering

\subsubsection{Contextual query}
To increase ease of use and reduce input to the memory assistant, queries from the user can be shortened using contextual awareness. As the device continuously tracks the context of the ongoing flow of the conversation, it enables the user to query the memory assistant with questions that build on this flow for a less disruptive interaction. For instance, if a user is saying the following sentence, \textit{"John teaches science, math and..."}, and wishes to recall the third subject that John teaches, with context awareness of the assistant, the user could directly query \textit{"What else?"} as opposed to having to formulate the full context-unaware query \textit{"What is the subject that John teaches other than science and math?"}.

\begin{figure*}
  \includegraphics[width=\linewidth]{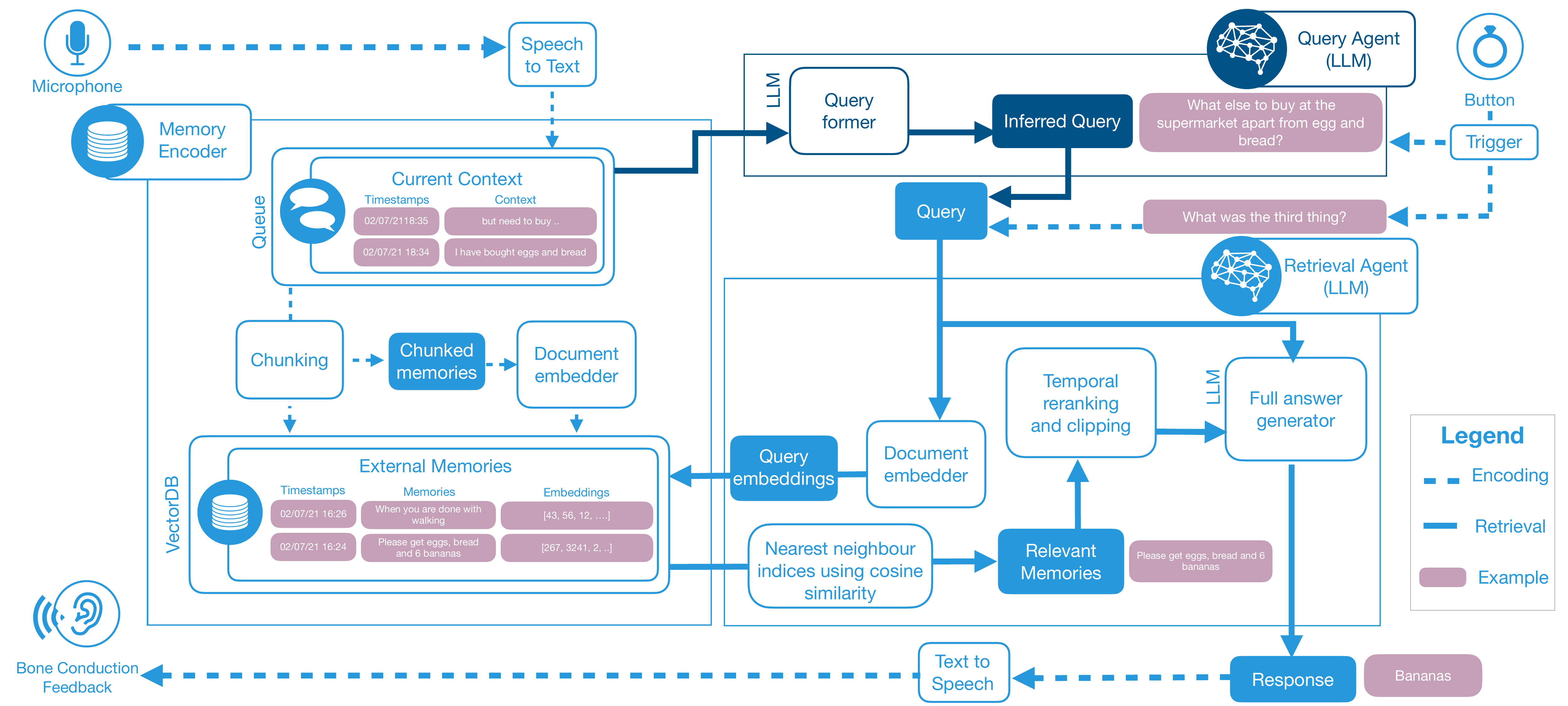}
  \caption{Detailed workflow of the components of Memoro: memory encoder, retrieval agent, and query agent. The memory encoder takes speech transcriptions and maintains the context and external memories. The query agent takes in the context and produces an inferred query. The retrieval agent takes a query and retrieves an answer from external memories.}
  \Description{Architecture of Memoro and its two interaction modes. It contains the system architecture of the memory assistant. It also showcases a comic strip of the two interaction modes: (1) Query Mode where the user can ask contextual questions (2) Queryless Mode where the user can request predictive assistance and skip query formation. For the modes, a person is speaking to the user (wearing the device) about another person whom they met at a conference and forgets the name. Subsequently, in query mode, the user uses a query to get the answer and speaks it out to the other person, and in queryless mode, gets an answer automatically and speaks it out to the other person integrating it into their speech. In both modes, responses are discreetly played back to the user using bone conduction.}
  \label{fig:detailed_workflow}
\end{figure*}

The contextual search is implemented using the following approach. When the user voices a natural language query to the memory assistant, the query and the \textbf{Current Context} containing the most recent conversation are combined to retrieve relevant external memories from the vector database. First, the vector embeddings for the query and current context, which are concatenated, are calculated using the same embeddings model used in the memory encoder. These vector embeddings are used to search for the most semantically similar external memories by comparing them to the stored embeddings of the \textbf{External Memories }which are pre-calculated during the encoding process. The comparison uses the established approximate K nearest neighbor search with cosine score as the similarity measure \cite{kushilevitz1998efficient}. The text transcriptions of the 10 most similar external memories constitute the relevant memories for the contextual search. The relevant memories are reordered based on ascending timestamps to form temporally linear memories and then clipped to the token limit (4096 tokens) of the large language model. The query, current context, and retrieved relevant memories are then combined, as described in Figure \ref{fig:detailed_workflow}, to form a prompt for the text generation language model. The prompt uses a combination of explicit and structured prompt engineering. Explicit prompts directly request the LLM to generate an answer to the user query from the relevant memories, while the structured aspect uses a template to guide the generation to a parse-able form. The prompt is designed to be able to search through relevant memories and generate the answer. The prompt can be found in Appendix \ref{appendixD_prompts}.

% The initial implementation of the retrieval agent used a technique with LLMs called Retrieval Augmented Generation \cite{lewis2021retrievalaugmented} (RAG). This method is able to look up documents for the answers using vector embeddings of them to ensure that the generated answer is from the documents. This approach presented an optimistic prospect in diminishing the hallucination of responses, albeit at the cost of real-time generation, as it was slower. As our user study, described in Section 4, involved limited memory length, there was no discernible difference in response quality when the aforementioned prompting technique was used, as it facilitated a faster generation of responses, a critical factor for usability, particularly in wearable technology applications. Hence, we used the prompting technique in our experiment however, we could switch to RAG when needed in a real-world usage context. 

\subsubsection{Concise Suggestions}
Once the answer has been retrieved using the above method, it is further post-processed to be more concise to minimize response duration and reduce output from the assistant. Searching through External Memories, rather than sifting through new information, allows for further conciseness \cite{DASGUPTA2021240}. For instance, \textit{"Her name is Sarah"} can be replaced with \textit{"Sarah"}. Therefore, the objective of this step is to eliminate any extraneous words such as connectives that do not address the question. Further, contextual compression could be used to remove any words that have already been retrieved by the user, either in the query or in the current conversational context. For instance, with the current context as \textit{"She is an engineer"} the query \textit{"What was her name and what is her specialization?"} and the generated answer \textit{"Her name is Emily and she works as a Software Engineer"} gets compressed to \textit{"Emily, Software"}. Addressing the query from the user, the answers can be shortened to specifically what is needed to complete the user's need. This is critical as language models tend to be more verbose as they are optimized for informativeness \cite{stahlberg2022conciseness}. The conciseness and redundancy removal are implemented by passing the query, current context, and the generated answer from the previous run to the retrieval agent with a template prompt that instructs the language model appropriately. The template prompt can be found in Appendix \ref{appendixD_prompts}.

\subsection{Query Agent}

In order to further streamline the interaction between the user and Memoro, we implemented an additional feature in the memory assistant that enables the user to receive on-demand predictive assistance without having to explicitly form a query, enabling the \textit{Queryless Mode}. This is facilitated by the user requesting the memory assistant to understand the ongoing flow of the conversation and infer their precise memory need. For example, if the user is already saying \textit{"He likes to play Settlers of Catan, Pandemic and ..."}, and then triggers the assistant, the query agent can predict the user query \textit{"What is the third board game he likes?"} allowing the user to skip query formation. To achieve this, we use a method that infers the query that the user is likely to ask based on a \textbf{Current Context} buffer, similar to the one implemented in the \textit{Query Mode}. The question inference leverages another iteration of prompting the language model to produce the query. The prompt can be found in Appendix \ref{appendixD_prompts}. The inferred query is then passed to the retrieval agent and the resulting concise answer is then presented to the user using text-to-speech synthesis. By implementing this feature, we aim to minimize the time spent in interactions during conversations, making Memoro more efficient and user-friendly.

\section{User Study}

To evaluate the interaction, usability, and experiences of users with Memoro, we conducted a within-subject study with N=20 participants and separated the two interaction modes for a detailed evaluation. In the study, the participants were introduced to fictional people and then engaged in a live conversation with the researcher about these fictional people. They experienced this in different conditions to evaluate the RQs.

\subsection{Tasks}
%To control the memory that is being interfaced with,
\subsubsection{Introductions to Fictional People}
We created four fictional people who were introduced to the participants, one for each condition. The introductions consisted of information-dense details such as the fictional persons' occupations, families, hobbies, and interests. The scripts are provided in the Appendix \ref{appendixA_fictional}. The introductions were played as audio {\color{black} with an image of the fictional person (generated using an online AI face generator\footnote{\url{https://thispersondoesnotexist.com/}}) displayed on screen} and were around 2 minutes long per person. The experiment was designed to make it very difficult to remember all these details. The introductions formed the External Memories for subsequent interactions with the memory assistant. No additional information was encoded into the External Memories during the conversation for a careful study of the interaction modes of the system. 
%[Add a figure to show how this looks like]

\subsubsection{Related Conversation}
To simulate scenarios where the participants would be in a real-time conversation and allow them to use the system, we engaged the users in an open-ended conversation consisting of scripted questions about the fictional people, with the researcher. For each fictional person, there were two general questions and four specific questions (see Appendix \ref{appendixB_scriptedqn} for more details). The researcher made sure to use the question set during the conversation. The responses from the participants were not scripted and they could choose when and how to interact with the system in the given condition at their discretion.

\subsection{Conditions}

The conditions were designed to elicit differences to technically and subjectively evaluate Memoro during the conversation. To address \textbf{RQ2}, we had a \textsc{No System} condition, where the participant engages in the task without the use of the system to compare and understand the effects on conversational quality and task load. In order to address \textbf{RQ3}, which was to determine the effect of contextual awareness and conciseness on the system's usability and user preferences, we set up a Baseline LLM system that is identical to the retrieval agent for question answering but does not use contextual awareness or conciseness. Therefore, participants needed to ask comprehensive questions in the Baseline condition and receive complete answers from the system. Overall, there were four conditions: 
\begin{itemize}
\item  \textsc{No System} which was the control condition 
\item \textsc{Baseline} LLM system with explicit query and raw, full-length answers
\item  \textsc{Query Mode} of Memoro with explicit contextual query and concise answers
\item  \textsc{Queryless Mode} of Memoro with no query and concise answers. 
\end{itemize}

In addition, technical evaluations were conducted to measure the system response accuracy and its conciseness. The interaction modes were separately analyzed for a detailed evaluation.

\subsection{Apparatus}
A web application showed the interface for playing the fictional introductions and was displayed on a 13" laptop. 
A Python program controlled the Baseline, Query, and Queryless Modes and was run on a separate laptop. As the Control did not involve any system and was based on free responses from participants, it did not require a separate laptop. For the three system conditions, participants wore a bone conduction headset \textit{Shokz OpenRun Pro} through which the participants interacted with the system.
To use the Baseline and the Query Mode during the experiment, participants held down a trigger key, on a wireless keyboard right in front of them, during which they voiced out their query. The query ended when they released the key. The Queryless Mode, as it did not require an explicit query, is invoked by a single press on the trigger key. The trigger key would be included as a ring button for mobile settings. All query inputs were using natural speech. The surveys were administered through an online platform.

\subsection{Measures}
We focused on evaluating the differences in the conditions in terms of response conciseness, accuracies and latency (\textbf{RQ1}), quality of conversations and task performance (\textbf{RQ2}), and user perceptions and experience (\textbf{RQ3}). %The full list of measures and questions are in the Supplementary Materials.

\subsubsection{Technical Evaluation of Interactions}
 %The interaction modes' technical aspects were evaluated objectively and subjectively. 
 {\color{black} The assistant's responses, users' queries, and interactions were automatically logged by the system.}
 The conciseness/verbosity of the assistant's responses was measured based on character count. The accuracy of the assistant's responses was manually evaluated after the study. {\color{black} The ground-truth of the responses is from the details of the fictional people used in the study. Each response was categorized: (1) Correct - if the response from the system was accurate to the query and context, (2) \textit{Don't Know} - if the required response was correctly identified to be not existing in the External Memories and the response was \textit{"I do not know the answer/Unknown"}. (3) Incorrect - if the response from the system was incorrect to the query and context, and (4) Speech Recognition Error - if the response was inaccurate due to a speech transcription error. The evaluation was conducted by two researchers who were independent of the data collection and blinded to the conditions.} The Query Time (how long users took to input their query for Baseline and Query Mode), processing time, and the total number of interactions were collected. We qualitatively analyze the overall systems' ability to respond to diverse queries asked by the participants in the study. 

\subsubsection{Quality of Conversations, Task Performance and Task Load}

For each condition, we measured the Quality of the Conversation as adapted from previous works measuring conversation quality~\cite{cai2023paraglassmenu}.
We measured six self-perceived aspects {\color{black} for the quality of the conversation rated using a 7-point Likert scale: listening to the conversational partner (`When the other person was speaking, I was always listening to them'), concentration on the conversation (`I was always concentrating on the conversation'), attention towards conversation partner (`When I was speaking, my attention was towards the other person'), eye contact (`When I was speaking I maintained eye contact.'), naturalness (`I acted naturally at all times during the conversation’), and feeling relaxed (`I felt relaxed during the conversation').} Following previous studies~\cite{cai2023paraglassmenu}, the perceived task load for engaging in the related conversation (with and without the systems) was collected using Raw NASA-TLX (RTLX~\cite{hart1988development}).
We also collected measures on their Task Performance/Recall Ability (with the systems if applicable) using a 100-point slider scale. The scale was chosen to match the NASA-TLX scale. It involved three aspects: confidence in recall ability, difficulty in recall, and recalled relevance. {\color{black} We used self-reported measures with questions designed to accurately reflect the hypotheses, and as there is a significant positive correlation between memory self-efficacy and memory performance~\cite{beaudoin2011memory}.} The full questionnaires can be found in Appendix \ref{appendixE_questionnaires}. 

\subsubsection{User Perceptions and Experience}
% Additionally, user experience and perceptions were measured (using a 7-point Likert scale). The collected measures were the rated Length of Responses (`I felt that the length of the answers were appropriate.'), Adaptiveness of the system (`I felt that the system adapted to my needs in the conversation.'), Interruption to Conversation (`The device manipulation by me interrupted the conversation.'), Helpfulness of Response (`The answers from the system were helpful.'), Usefulness (`The system would be useful in my everyday life.'), 
% Politeness (`I felt it was polite to use the system during the conversation.'), 
% Naturalness (`I acted naturally at all times while focusing on the researcher’s face and using the system.'), and ease of ignoring the device (`It was easy to ignore the fact that I was wearing the device.').
For each condition other than no system, we evaluated the System Usability using the System Usability Scale (SUS~\cite{brooke1996sus}).
Additionally, user experience and perceptions with respect to system usefulness and disruption caused were measured (using a 7-point Likert scale). The collected measures were the rated length of responses, adaptiveness of the system, interruption to conversation, helpfulness of response, usefulness of using the system, politeness of using the system, naturalness while using the system, and ease of ignoring the device. The full questionnaire can be found in Appendix \ref{appendixE_questionnaires}. Finally, we developed a post-study questionnaire where the preference rankings for all conditions and their reasons were collected. Open-ended questions were used to collect feedback on the overall experience in the study and suggestions for system improvements. {\color{black} The feedback was coded independently by two researchers (who were also independent of the data collection) and analyzed following Braun and Clarke~\cite{braun2006using} to generate initial themes. The researchers then reviewed the coded data and themes to come up with our final themes and analysis.} 

\subsection{Procedure}
\begin{figure*}[h]
    \centering
\includegraphics[width=\linewidth]{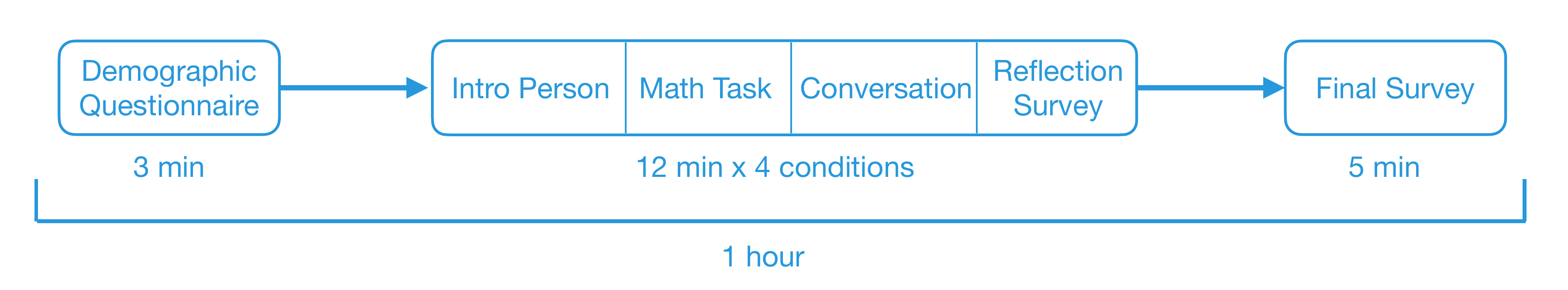}
    \caption{Procedure for the user study for each participant}
    \label{fig:procedure}
    \Description{Procedure of the user study. Starts with a demographic questionnaire, then intro to person, a math task and then the conversational recall, and ends with final survey. The full procedure is for one hour.}
\end{figure*}

Figure \ref{fig:procedure} summarises the study procedure. At the start of the study, participants were asked to fill in a demographic questionnaire (see Supplementary Materials). For Baseline, Query, and Queryless conditions, they were introduced to a fictional person while wearing the system. {\color{black} The order of the conditions was counterbalanced and the fictional introductions were presented in a randomized order.} After this, they engaged in a math task for distraction to refresh their short-term memory. Then, participants were asked to sit facing the researcher and engage in a conversation about the fictional person. Before the conversation (except in the "No System" condition), participants were given video instructions on how to use the system/mode in the respective condition. Participants familiarised themselves with the mode and practiced using it with an example. During the conversation, participants were able to use the system's features (except in the "No System" condition). Next, participants answered questionnaires about their experience (Section 4.4). 
At the end of the study, the participants answered a final questionnaire to rank their preferred condition, explained their ranking, and provided answers to open-ended questions on their experience using the system. The study took about one hour to complete {\color{black} and was set in a room within the laboratory}.  %The procedure is shown in Figure \ref{fig:procedure}.

\subsection{Participants}
Participants were recruited through email lists as well as snowball sampling and word-of-mouth. 20 participants took part in the study (9 male, 9 female, 2 non-binary, age range = 18 to 32, $age_{mean}$ = 23.4, $age_{SD}$ = 4.2 ). Participants were fluent or native English speakers with normal or corrected-to-normal hearing. Participants rated their listening memory between `Somewhat bad' (4), 'Neither good nor bad' (3), 'Somewhat good' (11), and 'Extremely good' (2). Additionally, the participants rated their frequency of experiencing tip-of-the-tongue moments in conversation as 'Never' (1), 'Sometimes' (13), 'About half the time' (4), and 'Most of the time' (2). The participants rated their frequency of using voice assistants as 'Not at all' (7), 'Once a month' (3), 'A few times' (4), 'Once a week' (3), 'More than once a week' (3).
The study received ethics approval from the university ethics review board, and participants gave written consent to take part in the study.

\section{Results}
We show the analysis from the user study of the systems' usability, technical evaluation, user perceptions and experience, and preferences.

%Figures {} and {} summarize the results (See Appendix for more details).

\subsection{Technical Evaluation}
A total of 392 interactions with the system were captured in the user study for all conditions: 102 for the Baseline, 150 for the Query Mode, and 140 for the Queryless Mode. Each interaction indicates a moment when the user requested memory assistance by using the button. We used these interactions for the technical evaluation.

\subsubsection{Conciseness and Processing Times}
The normality assumption for the response length data was not met according to the Shapiro-Wilk test ($p$<$.05$). Friedman tests (k=3) were conducted to determine if there were main effects in the conciseness. The test indicated significant differences between conditions (${\chi}^2$ = $135$, $p$<$.001$) in the response length from the system. The Wilcoxon signed-rank tests with Bonferroni correction in post-hoc showed that the Query Mode resulted in significantly shorter responses for the queries asked by the user as compared to the Baseline ($p$<$.001$), with an 85\% reduction in the mean number of characters from 115.4 to 16.6. The Queryless Mode has a response length similar to the Query mode. The average query time was also reduced by 15\% from 3.4 seconds for the Baseline to 2.9 seconds for the Query Mode ($p$=$.03$). The query time is not applicable for the Queryless Mode. The average processing time of the system for the Baseline and Query Mode was 1.4 seconds and 2.3 seconds for the Queryless Mode. The processing time reflects the time from the end of the query to the start of the audio feedback of the answer. Table~\ref{tab:freq} shows the detailed statistical results.

\begin{table*}[h]
  \caption{Average response length and average process time from the system, and average query time by the participant in the different conditions}
  \label{tab:freq}
  \begin{tabular}{lccc}
    \toprule
    Condition&Average Response Length (n chars)&Average Query Time (s)&Average Process Time (s)\\
    \midrule
    Baseline & 115.4$\pm$ 82.9 & 3.4$\pm$2.8 & 1.4$\pm$0.7\\
    Query Mode & 16.6$\pm$11.0 & 2.9$\pm$3.9 & 1.3$\pm$0.6  \\ 
    Queryless Mode & 21.1$\pm$11.8 & - & 2.3$\pm$0.8\\
  \bottomrule
\end{tabular}
\end{table*}

\subsubsection{Accuracy of responses generated by the System}
 Overall, the accuracy of the Baseline and Query Mode was 80.3\% and 84\% respectively. Notably, for 11.7\% and 6.0\% of interactions, the systems correctly determined that the question did not have an answer in the External Memories. Further, in the inaccurate responses, the participants could identify the inaccuracy and request the correct response with a different query. The Queryless Mode had an accuracy of 70.7\% and the drop was due to the Query Agent misinterpreting the context. For instance, during an interaction of P17, the Current Context contained \textit{"His favorite authors are Neil Gaiman and Ursula .."} and the inferred query was \textit{"What are William Thompson's hobbies and interests?"} which was incorrect as the participant was looking for the last name of \textit{Ursula}. However, we observed the response accuracy of the Queryless Mode was sufficient for a detailed evaluation. This was reflected by the final user preferences.
\begin{table*}[h]
  \caption{Accuracy of the responses generated from the system in the different conditions}
  \label{tab:freq}
  \begin{tabular}{lcccc}
    \toprule
    Condition&Correct (\%)&\textit{Don't Know} (\%)&Incorrect (\%)&Speech Recognition Error (\%)\\
    \midrule
    Baseline &  80.3&11.7&2.9&3.9 \\
    Query Mode & 84.0&6.0&6.7&3.3 \\ 
    Queryless Mode & 70.7&0.7&23.5&2.8 \\
  \bottomrule
\end{tabular}
\end{table*}

\subsubsection{Handling diverse queries from users}
\begin{figure*}[h]
    \centering\includegraphics[width=\linewidth]{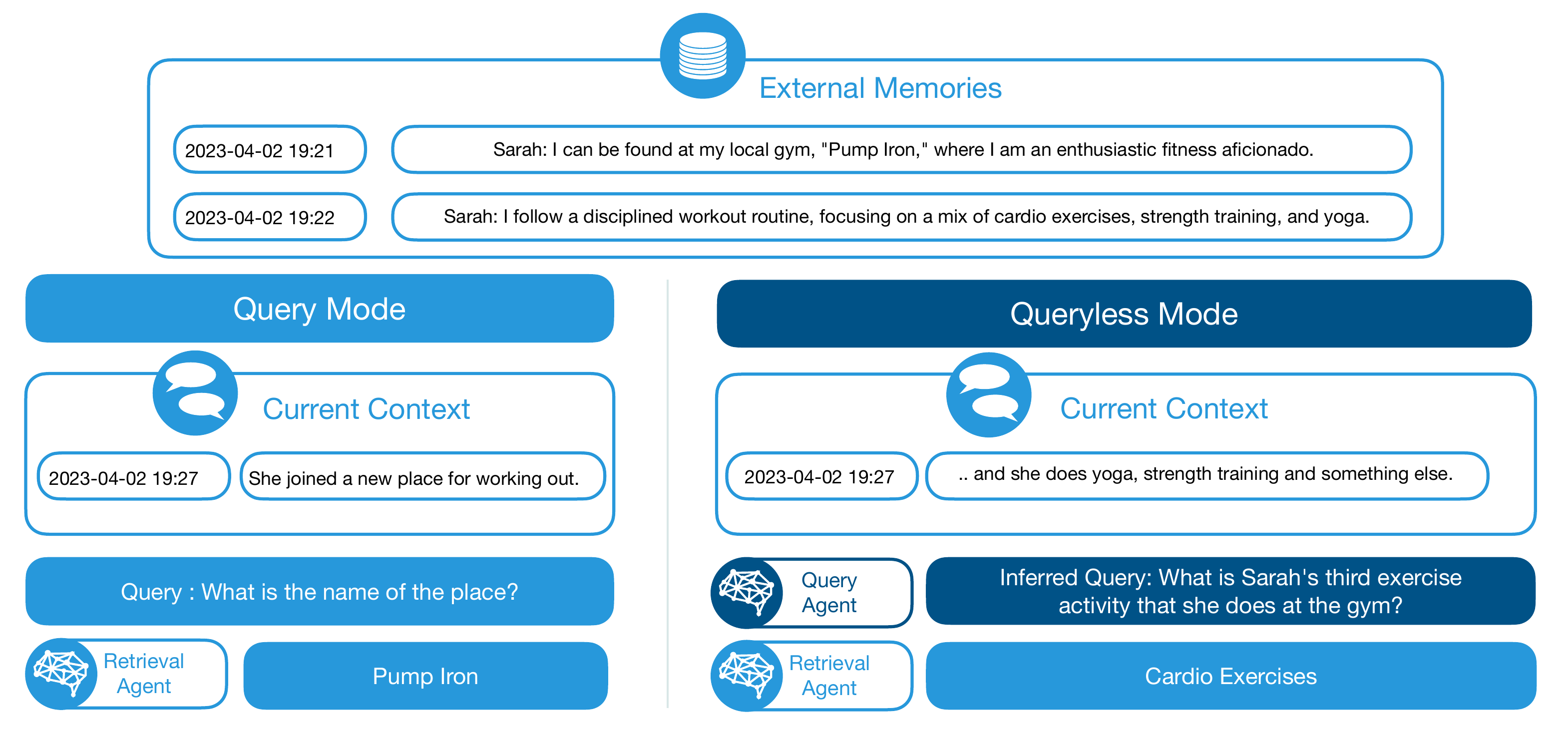}
    \caption{Example interactions by P3 and P19 show the Query Mode and the Queryless mode for the same memory respectively. The timestamps are changed for reporting.}
    \Description{Example interactions by P3 and P19 show the Query Mode and the Queryless mode for the same memory respectively. Both are recalling the gym routine of a fictional person. In query mode, the contextual query is "what is the name of the place?" and is sucessfully prompted with the answer "pump iron". In the queryless mode, the user is looking for the third exercise and asks Memoro for assistance, which correctly identifies the query and replies with "cardio exercises"}
\label{fig:example_interactions}
\end{figure*}

The usage of a large language model (LLM) allows the system to understand the intent of a user and enables natural language search beyond keyword matching, such as semantics. With sufficient information, it can predict the query by understanding the user's intent. The performance of the retrieval and query agents using LLMs are illustrated with the following examples of the interactions by two of the participants (P3, P19) in Figure \ref{fig:example_interactions}. In the first example (P3), the user opted to substitute the term \textit{`gym'} with the phrase \textit{`place for working out'}, and the retrieval agent comprehended the intention of the user. In the second example (P19), the query agent interpreted that the user was looking for the third activity and inferred a query for the retrieval agent, resulting in a successful interaction. More such examples can be found in Appendix \ref{appendixC_examples}.

\subsection{Conversation Quality, Task Performance, and Task Load between Conditions}

\subsubsection{Quality of Conversation} There were no significant differences in conversation quality between conditions for the measures of attention  (${\chi}^2$ = $3.63$, $p$=$.303$),  concentration (${\chi}^2$ = $7.21$, $p$=$.0655$), eye contact (${\chi}^2$ = $7.00$, $p$=$.0719$), and how relaxed they were during the conversation (${\chi}^2$ = $3.85$, $p$=$.278$). %The full statistical results are in the supplementary materials. 
The quality of the conversation was preserved and not reduced in any of the conditions. 
We found a significant difference in the naturalness of conversation between the conditions (${\chi}^2$ = $13.8$, $p$<$.01$). There were significant differences between the No System condition and the system conditions: No System-Baseline $p$<$.01$, No System-Query $p$<$.01$, No System-Queryless $p$<$.01$, No System $M$=$5.75$, $SD$=$1.41$, Baseline $M$=$4.30$, $SD$=$2.03$, Query $M$=$4.25$, $SD$=$1.94$, Queryless $M$=$4.55$, $SD$=$1.82$.

\subsubsection{Task Performance and Task Load}
\begin{figure*}[h]
  \centering
  \begin{subfigure}{0.22\linewidth}
    \includegraphics[width=\linewidth]{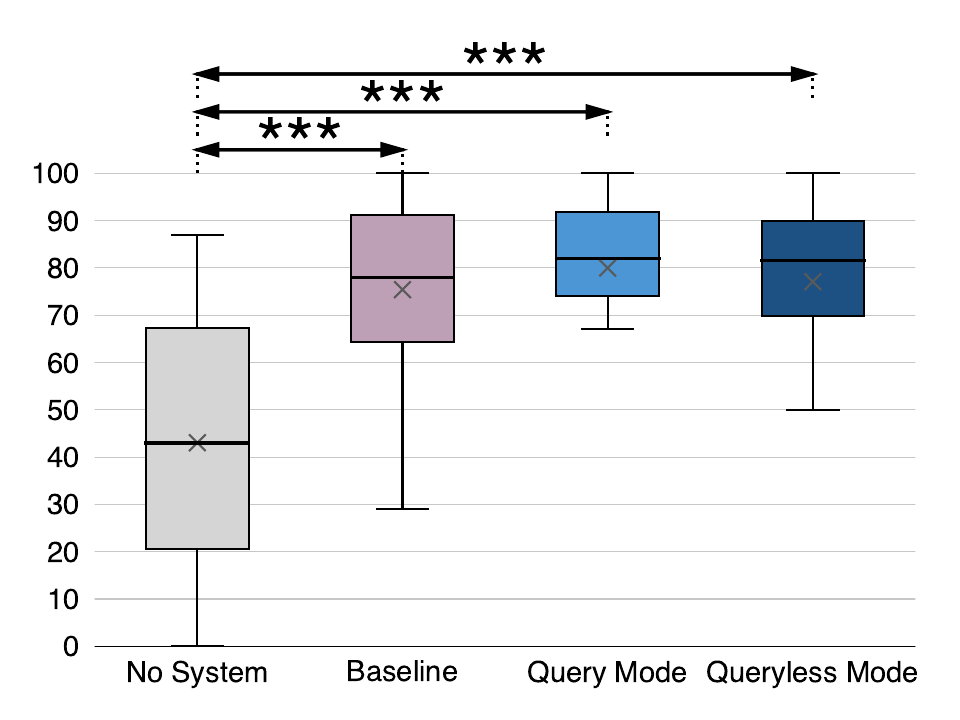}
    \caption{Confidence}
    \label{fig:subfig_a}
  \end{subfigure}
  \hfill
  \begin{subfigure}{0.22\linewidth}
    \includegraphics[width=\linewidth]{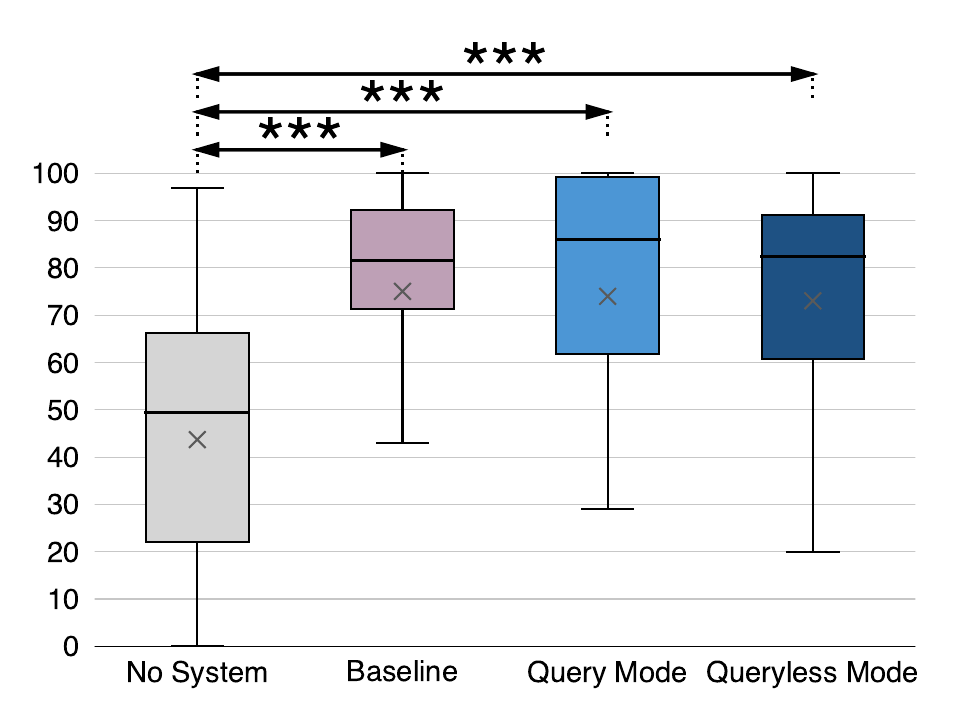}
    \caption{Recalled Relevance}
    \label{fig:subfig_c}
  \end{subfigure}
  \hfill
  \begin{subfigure}{0.22\linewidth}
    \includegraphics[width=\linewidth]{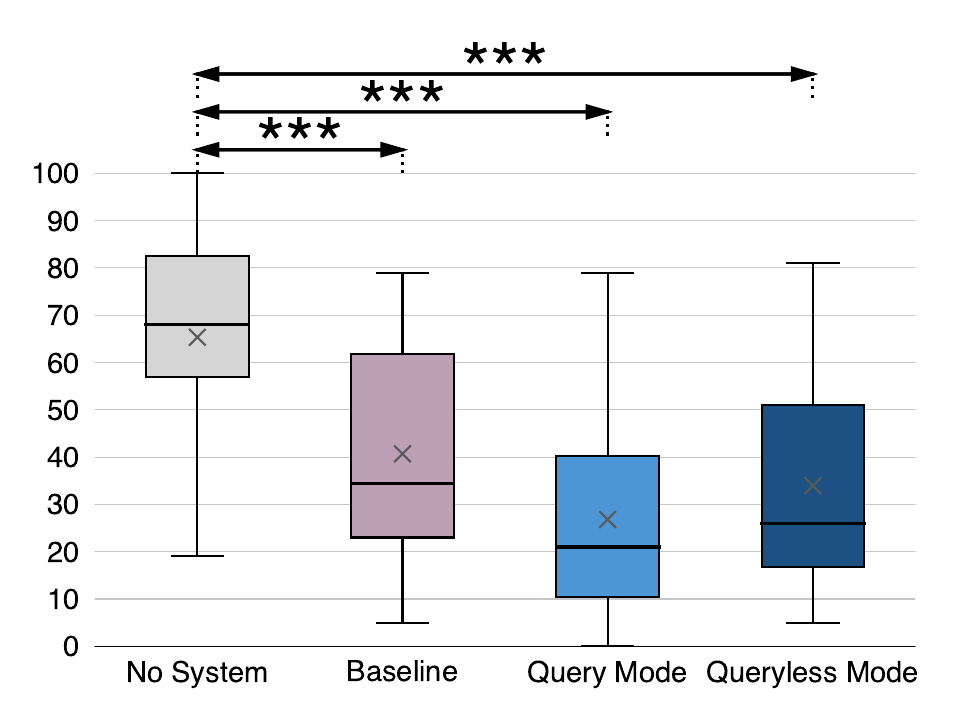}
    \caption{Difficulty}
    \label{fig:subfig_b}
  \end{subfigure}
  \hfill
  \begin{subfigure}{0.22\linewidth}
    \includegraphics[width=\linewidth]{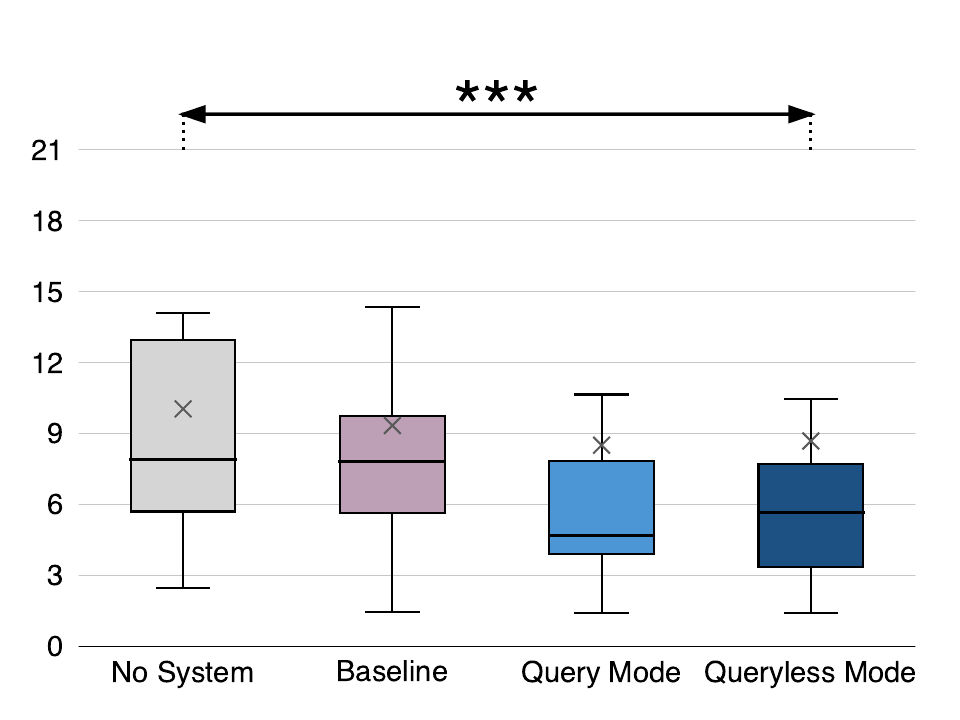}
    \caption{NASA TLX}
    \label{fig:subfig_d}
  \end{subfigure}
  \caption{Task Performance and task load : (a) Confidence in recalling, (b) Relevance of recalled information, (c) Perceived difficulty in recalling, and (d) Raw NASA TLX scores. ***: $p$<$.001$ }
  \Description{Boxplots of measures for Task Performance and task load : (a) Confidence in recalling, (b) Relevance of recalled information, (c) Perceived difficulty in recalling, and (d) Raw NASA TLX scores reflecting the values from the paper's evaluation section}
  \label{fig:task}
\end{figure*}

There was a significant difference in the confidence in recalling information between the conditions (${\chi}^2$ = $19.9$, $p$<$.001$, Figure~\ref{fig:task}a). Confidence in recalling was significantly higher in the system conditions compared to the No System condition and: No System-Baseline $p$<$.001$, No System-Query $p$<$.001$, No System-Queryless $p$<$.001$, No System $M$=$43.1$, $SD$=$26.9$, Baseline $M$=$75.4$, $SD$=$19.6$, Query $M$=$80.0$, $SD$=$17.8$, Queryless $M$=$77.1$, $SD$=$18.7$. 

There was a significant difference in the relevance of recalled information between the conditions (${\chi}^2$ = $18.5$, $p$<$.001$, Figure~\ref{fig:task}b). There were significantly higher relevance ratings for the system conditions compared to the No System condition: No System-Baseline $p$<$.001$, No System-Query $p$<$.001$, No System-Queryless $p$<$.001$, No System $M$=$43.7$, $SD$=$29.6$, Baseline $M$=$75.1$, $SD$=$26.3$, Query $M$=$74.0$, $SD$=$32.1$, Queryless $M$=$73.1$, $SD$=$24.6$. 

There was a significant difference in the difficulty in recalling information between the conditions (${\chi}^2$ = $12.1$, $p$<$.001$, Figure~\ref{fig:task}c). Participants found it significantly more difficult to recall information without the system compared to the system conditions: No System-Baseline $p$<$.001$, No System-Query $p$<$.001$, No System-Queryless $p$<$.001$, No System $M$=$65.4$, $SD$=$26.0$, Baseline $M$=$40.7$, $SD$=$24.4$, Query $M$=$26.8$, $SD$=$21.0$, Queryless $M$=$34.0$, $SD$=$24.8$. 

We found significant differences in task load (RTLX) scores between conditions (${\chi}^2$ = $12.0$, $p$<$.001$, Figure~\ref{fig:task}d). Post-hoc analysis showed a significant difference in RTLX between No System ($M$=$10.0$, $SD$=$7.06$) and the Queryless Mode ($M$=$8.68$, $SD$=$11.4$). Overall, the RTLX scores were generally lower in the system conditions compared to the No System condition: Baseline $M$=$9.34$, $SD$=$7.19$, Query $M$=$8.51$, $SD$=$9.93$. 

%The confidence in recalling information and completion both were significantly higher than control condition. 
%The NASA TLX was lower but not significantly different from the control condition.

%\subsubsection{Recall Ability}

\subsection{User Perceptions and Experience with Memoro}

\subsubsection{System Usability}
%A wearable memory assistant requires easy usability; using the system usability scale, we evaluated how the participant felt using each of the assistants independently. 

\begin{figure}[h]
    \centering
\includegraphics[width=0.9\linewidth]{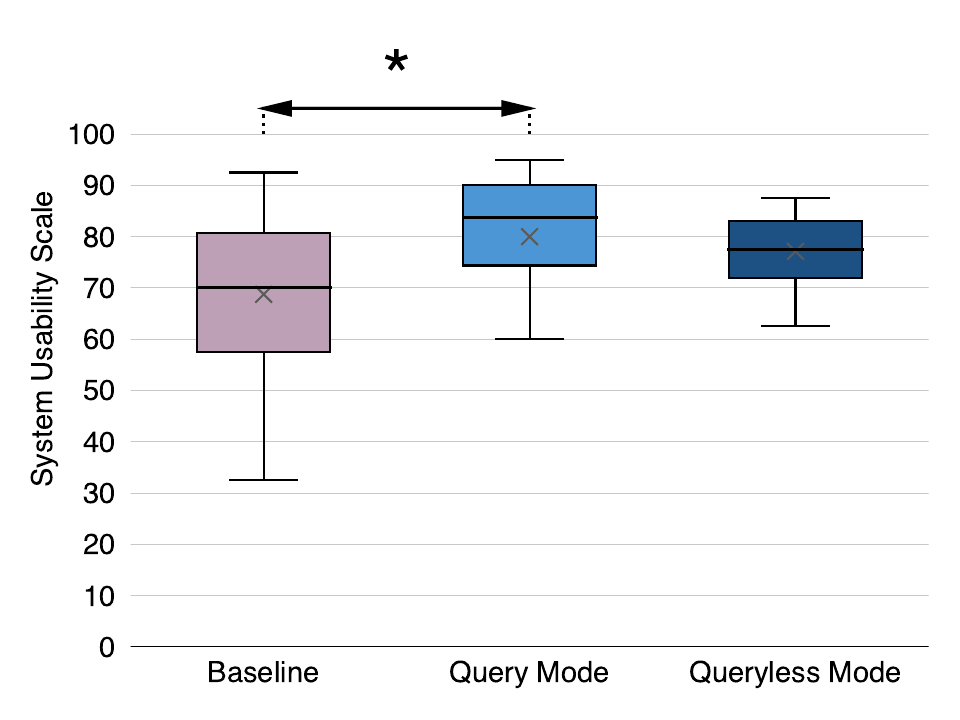}
    \caption{System Usability Scale (SUS) scores for the different assistants. *: $p$<$.05$}
    \label{fig:sus}
    \Description{Boxplot of System Usability Scale (SUS) scores for the different assistants and reflecting the values from the paper's evaluation section}
\end{figure}

The Query Mode of Memoro had the highest mean usability score of 80.0 ($SD$=$11.8$, Figure~\ref{fig:sus}). The Queryless Mode had a usability score of 77.1 ($SD$=$8.1$) and the Baseline had the lowest usability score of 68.75 ($SD$=$15.15$). Since the data was normally distributed according to the Shapiro-Wilk test ($p$>$.05$), a repeated measures ANOVA showed a main effect of the systems on the usability score ($F_{(2,38)}$=$5.053$, $p$=$.011$). A Tukey HSD post-hoc test showed a significant difference ($p$=$.015$) between the usability of Baseline and Query Mode. 

The normality assumption for the rating data was not met according to the Shapiro-Wilk test ($p$<$.05$). Friedman tests (k=3) were conducted to determine if there were main effects of the system conditions on the measures. Wilcoxon signed-rank tests with Bonferroni correction were used for post-hoc analysis when effects were found. 

\subsubsection{Rated Length of Responses}
The Friedman test indicated significant differences between conditions (${\chi}^2$ = $26$, $p$<$.01$) in the rated appropriateness of the response lengths (Figure~\ref{fig:perceptions}a). The Query ($M$=$5.55$, $SD$=$1.05$) and Queryless ($M$=$5.45$, $SD$=$1.19$) Modes had significantly higher ratings in length appropriateness (Query-Baseline:$p$<$.01$, Queryless-Baseline: $p$<$.01$) compared to the Baseline ($M$=$2.80$, $SD$=$1.47$). There were no significant differences between Query and Queryless Modes ($p$=$.685$).

\subsubsection{Adaptiveness of System}
Adaptiveness is defined as how closely the system is able to monitor the current context of the conversation with respect to the user perception. There was a significant difference between conditions (${\chi}^2$ = $11.7$, $p$<$.01$) in the rated adaptiveness to the conversation (Figure~\ref{fig:perceptions}b). The Query  ($M$=$5.35$, $SD$=$1.31$) and Queryless ($M$=$5.10$, $SD$=$1.07$) Modes had significantly higher ratings in adaptiveness (Query-Baseline:$p$<$.01$, Queryless-Baseline: $p$<$.01$) compared to the Baseline ($M$=$3.40$, $SD$=$1.85$). There were no significant differences between Query and Queryless Modes ($p$=$.448$).

\subsubsection{Device Interruption}
The Friedman test showed a main effect of conditions on device interruption to the conversation (${\chi}^2$ = $7.43$, $p$=$.0243$, Figure~\ref{fig:perceptions}c). However, post-hoc analysis did not indicate any significant differences between the conditions: Baseline $M$=$ 5.55$, $SD$=$1.28$, Query $M$=$4.40$, $SD$=$1.60$, Queryless $M$=$4.65$, $SD$=$1.53$, Query-Baseline:$p$=$.0173$, Queryless-Baseline: $p$=$.0362$, Queryless-Query: $p$=$.498$.

\subsubsection{Helpfulness and Usefulness}
\begin{figure*}[htbp]
  \centering
  \begin{subfigure}{0.22\linewidth}
    \includegraphics[width=\linewidth]{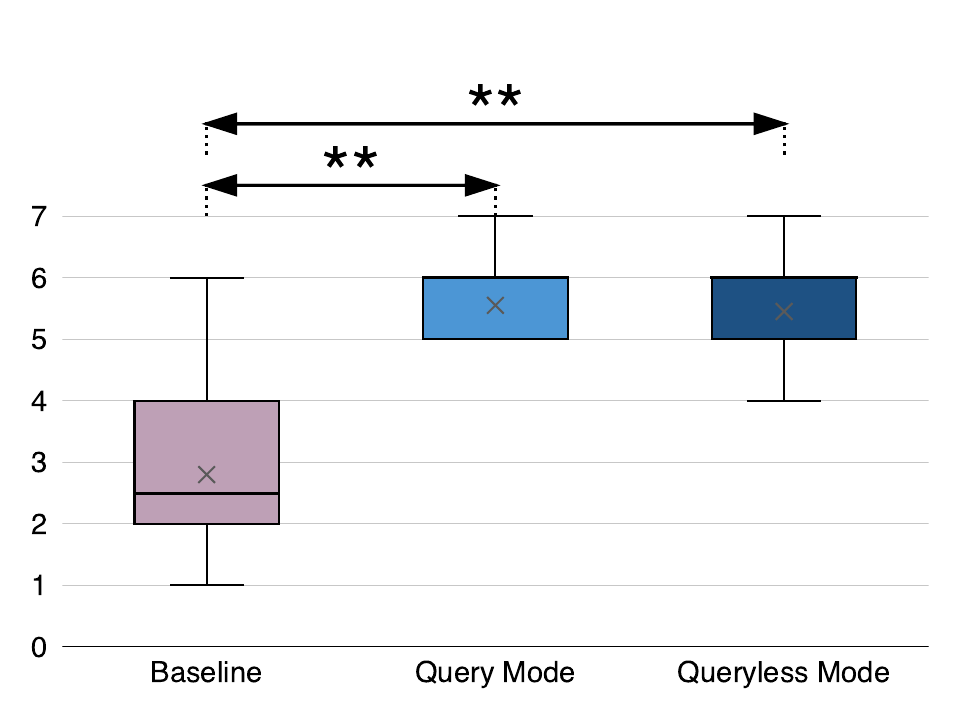}
    \caption{Appropriate Length}
    \label{fig:subfig_a}
  \end{subfigure}
  \hfill
  \begin{subfigure}{0.22\linewidth}
    \includegraphics[width=\linewidth]{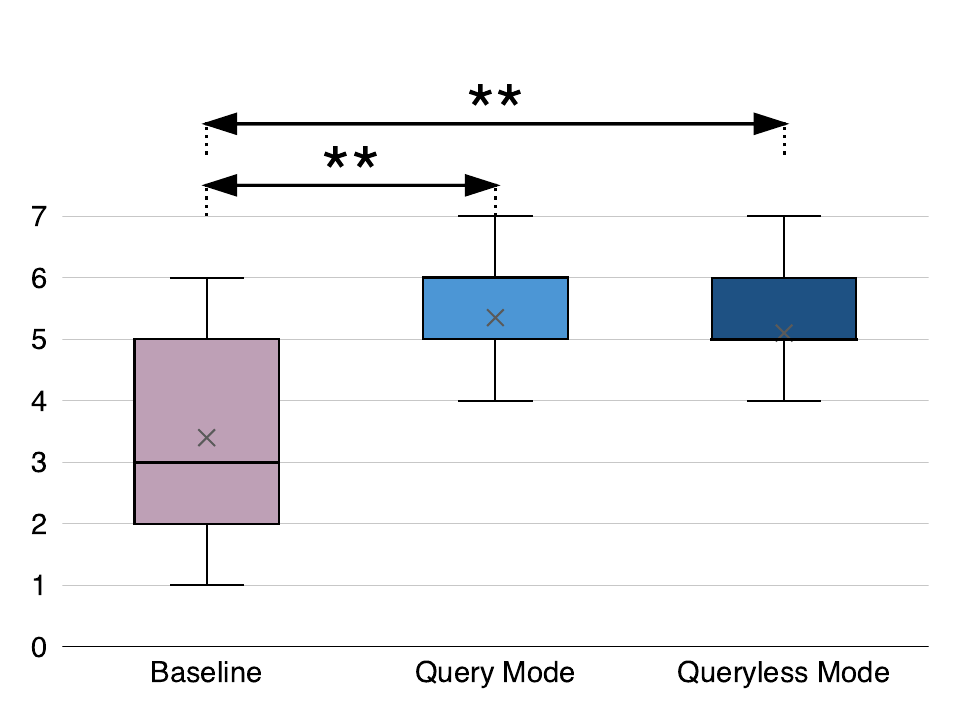}
    \caption{Adaptibility}
    \label{fig:subfig_c}
  \end{subfigure}
  \hfill
  \begin{subfigure}{0.22\linewidth}
    \includegraphics[width=\linewidth]{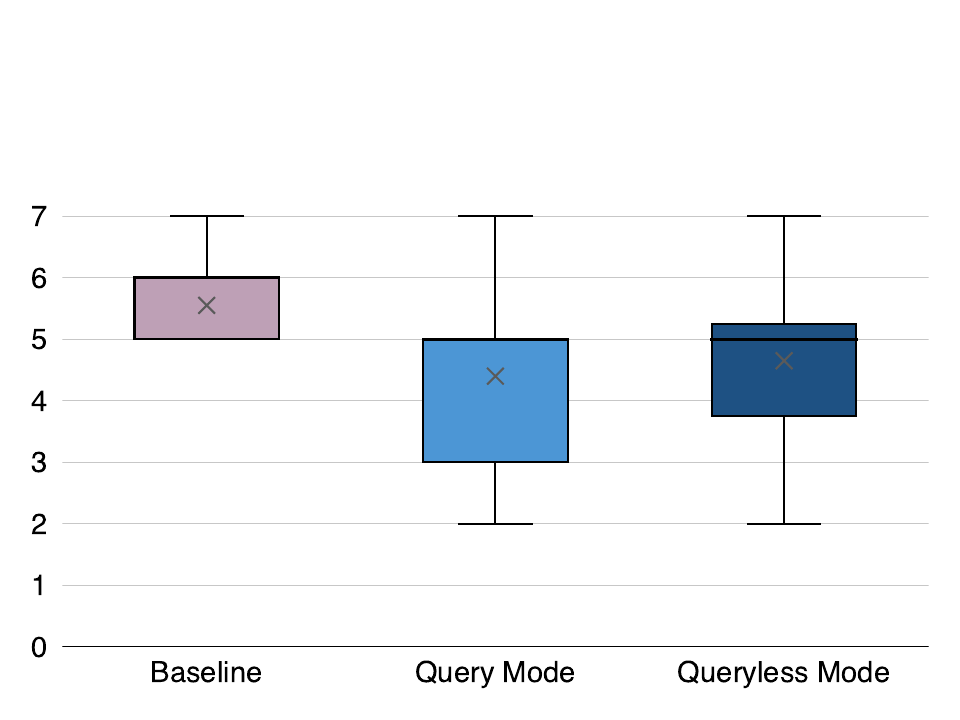}
    \caption{Device Interruption}
    \label{fig:subfig_b}
  \end{subfigure}
  \hfill
  \begin{subfigure}{0.22\linewidth}
    \includegraphics[width=\linewidth]{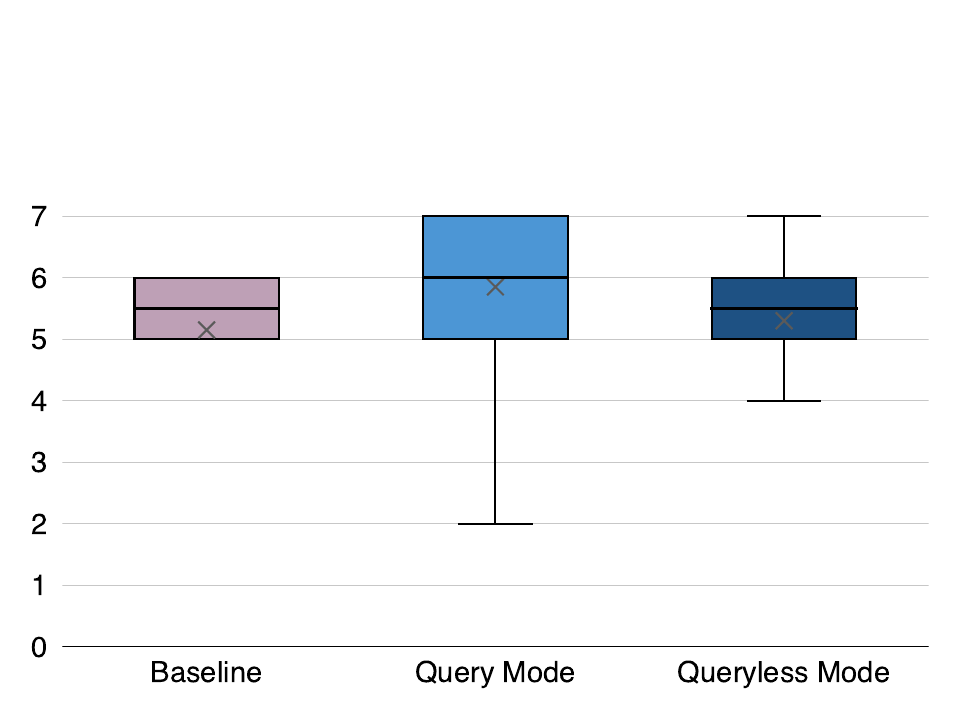}
    \caption{Helpfulness}
    \label{fig:subfig_d}
  \end{subfigure}
  \caption{User perceptions and experience of the different assistants: (a) Appropriateness of response length, (b) Adaptability, (c) Device interruption and (d) Helpfulness.  **: $p$<$.01$}
  \Description{Boxplots of User perceptions and experience of the different assistants: (a) Appropriateness of response length, (b) Adaptability, (c) Device interruption and (d) Helpfulness compared across the conditions and reflecting the values from the paper's evaluation section.}
  \label{fig:perceptions}
\end{figure*}

There was no significant difference in the conditions in terms of helpfulness (Figure~\ref{fig:perceptions}d): ${\chi}^2$ = $4.25$, $p$=$.119$, Baseline $M$=$5.15$, $SD$=$1.18$, Query $M$=$5.85$, $SD$=$1.18$, Queryless $M$=$5.30$, $SD$=$1.30$.
There was a significant difference in usefulness between the conditions (${\chi}^2$ = $11.9$, $p$<$.01$). Post-hoc analysis showed a significantly higher rated usefulness ($p$<$.01$) for Query Mode ($M$=$5.50$, $SD$=$1.36$) compared to the Baseline ($M$=$4.30$, $SD$=$1.53$). No significant differences were found between Baseline and Queryless Mode ($M$=$5.05$, $SD$=$1.43$, $p$=$.0358$), and Query and Queryless Modes ($p$=$.233$).
%This is more like discussion point: This indicates that the Concise and Queryless systems could maintain helpfulness while reducing the interaction time and length of responses.

\subsubsection{Politeness, Naturalness, Ease of Ignoring Device}
The Friedman test showed a significant difference in reported politeness of using the device in the conditions (${\chi}^2$ = $8.10$, $p$=$.0174$). Post-hoc analysis showed a significant difference in politeness ($p$=$.0144$) between the Baseline ($M$=$2.90$, $SD$=$1.37$) and Query Mode ($M$=$3.70$, $SD$=$1.45$). No significant differences were found between Baseline and Queryless Mode ($M$=$3.65$, $SD$=$1.35$, $p$=$.0420$), and Query and Queryless Modes ($p$=$.897$).

We found no significant difference in the conditions in how natural users acted (self-reported): ${\chi}^2$ = $3.30$, $p$=$.192$, Baseline $M$=$3.35$, $SD$=$1.57$, Query Mode $M$=$3.45$, $SD$=$1.67$, Queryless Mode $M$=$4.05$, $SD$=$1.47$. There was also no significant difference in the conditions in how easy it was for the participant to ignore that they were wearing the device: ${\chi}^2$ = $.128$, $p$=$.938$, Baseline $M$=$4.00$, $SD$=$2.03$, Query Mode $M$=$4.15$, $SD$=$1.63$, Queryless Mode $M$=$4.10$, $SD$=$1.59$.

\subsubsection{User Preferences and Qualitative Feedback}
\begin{figure}
    \centering
    \includegraphics[width=0.9\linewidth]{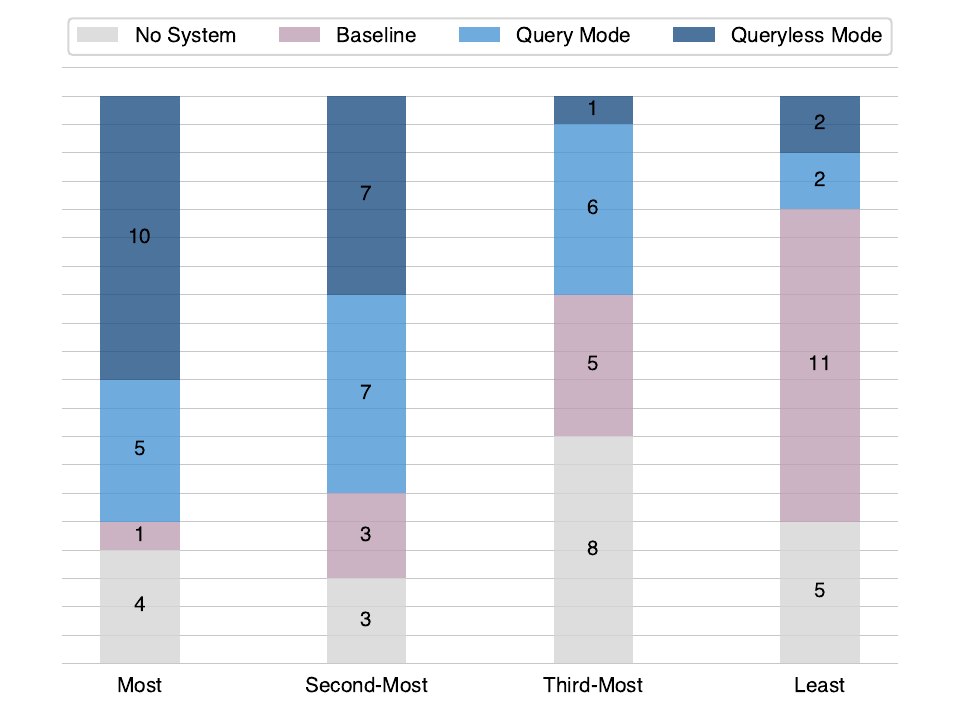}
    \caption{User preferences between conditions. The plot shows the number of participants who preferred which condition the most, the second-most, the third-most and the least.}
    \Description{Lineplot of the user preferences between conditions. The plot shows the number of participants who preferred which condition the most, the second-most, the third-most and the least. Each color is a different condition.}
    \label{fig:pref}
\end{figure}
The preference rankings are shown in Figure~\ref{fig:pref}. 10 of 20 participants preferred the Queryless Mode the most, and 11 of 20 preferred the Baseline the least. 

Participants felt that the Queryless Mode felt the \textit{``most seamless''} (P19) and that it was \textit{``very nice and barely noticeable''} (P14). They explained that they preferred it the most because it \textit{``preempts context''} (P17), it \textit{``required the least amount of effort''} and \textit{``anticipated''} their needs (P12) and questions (P10).
Participants also reasoned that it was the \textit{``best in terms of real-life usage, mainly because using it interrupted conversation the least''} (P4) and it \textit{``made the conversation less awkward''} (P7). P16 explained that Queryless mode was preferred to the Baseline and Query mode as \textit{``it seems a bit difficult and rude to ask question to the device, while I am still in conversation with the person''}.
Although it can be useful, P15 felt that more practice is needed to get used to using it: \textit{``...given some practice, I think the first questionless one has potential to be super useful with some practice. I just need to know when to hit the button for best results.''}

A few participants preferred the Query Mode over the Queryless Mode. P20 explained that \textit{``[The Query Mode] is slightly higher [ranked] because I could ask a question and felt the other person knew that I was consulting someone else for the answer which made it more slightly OK than [the Queryless Mode]''}. It \textit{``felt more appropriate/polite to use''} (P15) and it was the \textit{``most easily integrated into the conversation''} (P6). In some cases, users felt that the Query Mode had higher accuracy (P1, P2, P3, P4, P7) and \textit{``was better at answering''} (P5). 

Most participants (16 out of 20) preferred at least one of the system conditions over the No System condition. The users who preferred having No System explained the systems as \textit{``clunky''} (P17), or it depended on the task (P18); %P6 felt that the systems were not needed for the related conversation as they saw no need for remembering the information and felt that the conversation was \textit{``just small talk''}. 
P5 explained \textit{``I prefer natural conversation more which was easier without the assistant.''} 

Many participants felt that the Baseline was too lengthy (P2, P6, P7) and \textit{``to the point it was a little distracting''} (P5). P8 mentioned \textit{``...[it] went on for a long time and there wasn't a way to get it to stop or ask it to get to the point without waiting and stalling the conversation. I'd rather just move on and just leave it than have to wait unless it's REALLY important.''}
A few users preferred having No System over Baseline because it \textit{``broke the conversation flow too much to be preferred over no system.''} (P11) and \textit{``[The Baseline] is ranked 4th because it provided redundant answers and didn't actually adapt to the conversation. I felt like it wasn't as useful as just having to remember information off the top of my head.''} (P4). P10 liked Queryless Mode the most but ranked No System over Query Mode and Baseline: \textit{``But if I have to explicitly ask it questions, I would prefer to just rely on my memory''}.

\section{Discussion}

We discuss the study findings and to what extent they address the RQs.

\subsection{Integrating LLMs in Wearable Memory Augmentation}
Discussing RQ1: ``How can we design a seamless wearable memory assistant using LLMs to reduce disruption to the primary task with minimal and effective input and output?''
As recent advances in LLMs lead to improved capabilities in natural language processing tasks such as question answering and summarization, we found that using them in a wearable can facilitate a concise and seamless interface. It can be helpful to users for memory retrieval as all system conditions including baseline had a ``helpfulness'' ratings above 5.15 of 7. We found that our approach of introducing minimal output from Memoro using LLMs reduced perceived disruption/interruption (Baseline $M$=$5.55$, Query $M$=$4.40$, and Queryless $M$=$4.65$, out of 7) while preserving their helpfulness. The use LLMs in semantic search of memories also showed that they could improve flexibility in querying by allowing users to use synonyms or alternate phrasings. This contributed to the highly rated adaptiveness of Memoro for both modes (Query $M$=$5.35$ and Queryless $M$=$5.10$, out of 7) to the conversation and significantly higher ratings compared to Baseline. Through the Queryless Mode, we also demonstrate that LLMs can handle understanding user intentions in memory retrieval tasks during a conversation for minimal input. The conciseness of output was significantly improved methodologically with an $85\%$ reduction in answer length, and users rated them as having improved appropriateness of response length as compared to the Baseline condition. 
Overall, through the two modes of interaction of Memoro, we show a method of using LLMs for a concise interface in memory retrieval by providing flexibility in queries, understanding conversational context, and improving conciseness in responses.

\subsection{Impact of using Memoro in a Primary Task}
Discussing RQ2: ``What are the effects of using the memory augmentation system during the primary task of a real-time
conversation across metrics such as quality of conversation, performance, and task load?''
The emphasis on minimal disruption as being a core design principle for Memoro was to enable seamless interactions by users with their external memories while being preoccupied with a primary task, such as a conversation. Further discussing RQ2, in our study with social interactions, we validated that using Memoro did not affect the conversational quality in terms of attention, concentration, eye contact, or how relaxed they were as compared to when they used no system. The only aspect that was affected was that the conversations felt more natural with the No System condition compared to the system conditions. Along with this, participants showed a significant increase in recall confidence, a significant decrease in difficulty in recalling answers, and a significant increase in the amount of relevant information recalled during both modes of Memoro. The use of Queryless mode also resulted in a significant decrease in task load compared to the No System condition, making the conversation task cognitively easier for the user.

\subsection{Usability, Preferences, and Experiences}
Discussing RQ3: ``How do context awareness and conciseness affect the system’s usability, user perceptions, and experience?''
Overall, on evaluating the usability of Memoro, we find that the highest mean SUS score is for the Query Mode (80.0), followed by Queryless Mode (77.1). By adding contextual awareness and conciseness to the responses, there was a significant improvement in the usability from the Baseline LLM (68.8) condition. The SUS score of 80.0 lies in between the good and excellent range and is considered acceptable as it is well above the average score of 68 \cite{brooke1996sus}. This was further reflected in user preferences where 19 out of 20 participants rated a mode of Memoro over the Baseline and participants also mentioned that they would rather have no system and rely on their own memory over Baseline mode (Section 5.5). 
When analyzing the SUS scores for Memoro, previous work on comparing interfaces for Internet-of-Things (IoT) device manipulation during conversations showed that voice interfaces only achieved the mean SUS score of 70.88~\cite{cai2023paraglassmenu} compared to a visual head-mounted display with a score of 83.
One of the reasons for the longstanding issues with voice interfaces~\cite{Chan2020} is the accuracy of speech-to-text recognition. Although the recognition tool for Memoro and the previous study's tested voice interface was the same: Google Speech-to-Text API (Google Assistant), Memoro received higher usability scores and this might have been due to the use of LLMs to ``offset" the inaccuracy of the speech-to-text. These findings indicate that an important consideration in designing wearable memory retrieval assistants is to enable the users to ask brief questions and get concise and to-the-point answers. Our findings can inform further work on integrating LLMs into the wearable context.

While Query mode was the most usable and the most accurate (Section 5.2.2), Queryless Mode was the most preferred condition among the participants (10 out of 20). From the NASA RTLX scores, an explanation for these preferences could be the significant decrease in task load when using the Queryless Mode compared to using no system in the task. 
In addition to its good usability and accuracy, we argue that there is value in the Query mode too as it had significantly higher rated usefulness (Section 5.3.4) and felt more polite to use compared to the Baseline (Section 5.3.5). 
Further, on examining the participants who preferred `No System' over any of the other conditions, hence preferring no memory assistance (P7, P16, P18, P19), we found that two of them (P7, P16) rated their auditory memory as `Extremely Good'. They were the only two participants with that rating in the study. The other two (P18, P19) indicated that they have never used voice assistants in their daily life. This aligns with previous studies \cite{tabassum2019} that people perceived increased benefits of voice assistants if they had used them before. These preferences indicate the need for more research into the influence of these factors in the design of wearable memory assistance.

\section{Limitations and Future Work}
While we show how Memoro was preferred by a majority of participants and was considered acceptable usability, we discuss the following limitations in the design and study of the wearable memory assistant. 

\subsection{{\color{black} Technical Aspects}}
Firstly, the encoding of external memory is based on timestamps and direct transcription of the recording of audio, inspired by existing lifelogging tools \cite{Vemuri2004, hayes2004}, and as the focus of the study was to explore minimally disruptive memory retrieval during a primary task. Integrating more information such as location, non-verbal gestures, facial expressions, and recognition of the conversation partner during memory encoding, can significantly advance the memory assistant by understanding more of the user's context~\cite{sarawgi2020multimodal,chan2022augmenting}. {{\color{black} The location (from GPS sensor) and conversation partner information can assist in filtering older memories for accurate retrieval. Non-verbal gestures can give insightful information on body language such as low engagement or heightened nervousness which can increase the importance of the memories encoded during that period. The importance could be further modulated by users explicitly. These features can enable diverse queries of the form} \textit{``Who did I meet in the cafeteria yesterday?"} or \textit{``What was the name of the person Ann spoke to me about 2 days ago?"}. {\color{black} Further, implicit prompting based on disfluencies in speech, and accelerometer-based gestures can reduce input effort and time by having users perform subtle hand gestures instead of clicking the trigger button. Implicit prompting can lead to studies understanding how short the query needs to be for a conversation to seem "uninterrupted" from an external perspective.}

Secondly, the use of LLMs in information retrieval can lead to hallucinated answers that do not exist in the database. The memories can also contain conflicting information which can lead to incorrectly generated suggestions. While tackling hallucinations in LLMs is an ongoing challenge, future work can address these concerns with a more sophisticated knowledge graph of the user's memories.

Thirdly, while we look at discreet audio feedback from the system to maintain eye gaze and reduce distraction during conversations, {\color{black} we acknowledge that there is a chance of the masking of the conversation with sound coming from the voice interface and voicing queries (for Query Mode) might disrupt the conversation. The timing for receiving the audio feedback is determined by the user, as such, users can choose to trigger retrieval during breaks between sentences (for Query and Queryless modes) or potentially mask queries within the conversation such as by rephrasing the conversational partners' questions (for Query mode).} Some users may prefer an Optical head-mounted display (OHMD) for visual feedback. For users who prefer OHMD, a similar assistant with visual answers could be given where the text-to-speech of response can be skipped. A study evaluating the pros and cons of audio-based versus heads-up display-based interaction in memory assistance would be an interesting next step.

\subsection{{\color{black} Study Design and Population}}
Next, the participants were from a group from the local community who may be more accustomed to such technology as voice assistants. {{\color{black} The experiment also was situated in a lab setting for a controlled study. Longitudinal and in-the-wild studies situated in natural settings with a geographically diverse user group while enabling both retrieval modes simultaneously are needed to understand the usefulness and applicability of Memoro in daily life outside of laboratory-based social interactions. Relatedly, longitudinal studies can employ text similarity algorithms to aid in the objective measurement of the recall ability of users.} Similarly, future directions include field studies with a specific subpopulation with a higher frequency of memory assistance needs, such as the elderly, where such a system could be more useful. An example is the tip-of-the-tongue (TOT) scenario \cite{BROWN1966}, {\color{black}forgetting of certain words}, which commonly occurs in older adults and increases with Aphasia. There can be an exploration of other forms of information presentation where, instead of giving direct answers, the system would give users episodic or semantic clues and optional answers~\cite{ha2023you}, or answers in voices of people you admire or are familiar with~\cite{chan2021kinvoices}.%https://dl.acm.org/doi/10.1145/3544549.3585670
% \cite{ha2023you}

\subsection{{\color{black} Privacy and Social Acceptability}}
Finally, it is important to consider legal, ethical, {\color{black} privacy and social acceptability} issues in deploying memory assistants that record audio from everyday conversations. 
{\color{black} Ensuring data security for pervasive memory augmentation systems is critical beyond ensuring encrypted data storage~\cite{davies2015security}. 
As research in psychology~\cite{anderson1994remembering} shows how we are prone to the simultaneous reinforcement of recovered memories and attenuation of unrecovered memories, memory augmentation interfaces can contribute to unintended altering and manipulation of captured memories through its process of retrieval. 
With the increase in the subtleness of wearables with recording capabilities~\cite{iqbal2023adopting}, future memory augmentation systems need to implement concrete and transparent methods, such as speaker verification \cite{reynolds2000speaker}, to manage permissions of recording. As this system is geared for daily use, the privacy of bystanders in the vicinity needs to also be accounted for.}
Further, in some states and countries, recording other people without their knowledge is illegal.
While this work assumes consent for recording from all parties involved, possible methods to address privacy controls in natural settings may be to record synthesized notes, rather than direct transcriptions, to require opt-in or enable opt-out{\color{black}, and be able to selectively erase data on request.} 
{\color{black} 

Social acceptability of lifelogging devices can be situational~\cite{davies2015security}, where certain contexts such as during sports and meetings can be more permissive to it as compared to intimate conversations and in healthcare settings. Cultural beliefs and user stereotypes could also shape the social perceptions of wearables and user-worn recording devices~\cite{schwind2020anticipated,kelly2018wearer,hsieh2016designing}. Bystander considerations also play a role in social acceptability where interactions that provide an explanation \cite{willamson2011} are likely to be better acceptable than fully hidden interactions. Future research efforts should focus on designing strategies to improve social acceptability, possibly following guidelines in social acceptability research in HCI~\cite{koelle2020social}.
}

Such issues are not dealt with in the current design of the Memoro system and are important areas of future research. Overall, we are cautiously optimistic, based on this first experiment, that systems like Memoro may one day assist people who can use help with information and memory retrieval.

\section{Conclusion}
We implemented and studied a concise user interface for an audio-based wearable memory assistant, Memoro, by using LLMs to make it minimally disruptive. By comparing it with a control condition without any system during a real-time conversation task, we evaluate how Memoro increases recall confidence and reduces task load while preserving conversational quality. By comparing it with a baseline LLM, we demonstrate how the disruption caused by using a wearable for memory augmentation can be significantly decreased by adding contextual awareness and conciseness to the suggestions. We further show how a majority of the participants prefer on-demand predictive assistance (Queryless Mode) over explicitly voiced queries (Query Mode) for real-time memory retrieval using wearables. Finally, we engage in open-ended feedback to understand users' preferences, experiences, and reservations for such a system and its interaction modes. Through this work, we contribute towards integrating LLMs into wearables for real-time memory augmentation and information retrieval, assessing their potential for minimal disruption and adaptability.

\bibliographystyle{ACM-Reference-Format}
\bibliography{sample-base}
\newpage
\appendix
\section{Fictional People}
\label{appendixA_fictional}
The introductions of the four fictional people were as follows:

\textbf{William}
``My name is William Thompson, and I am a 42-year-old software engineer residing in the bustling city of Austin, Texas. As a graduate of the University of Texas, I specialize in developing cutting-edge mobile applications for the renowned tech firm, VirtuTech Solutions, where I have worked for the past 15 years. Despite the high-pressure nature of my job, I am known for my calm demeanor and exceptional problem-solving skills, which have contributed to my professional success. I have created two major-selling apps, BuzzPal and FoodMingle.
Living in a modern, two-bedroom apartment in the heart of the city, I enjoy the convenience of urban life while also appreciating the serenity of my well-maintained complex. My living space is equipped with the latest smart home technology, reflecting my keen interest in gadgets and innovation.
I am a proud father of two energetic children, 12-year-old Emily, a budding violinist, and 9-year-old Ethan, who has a passion for soccer. Emily and Ethan attend a local Montessori school, and I share parenting responsibilities with my wife, Lauren, a high school teacher who specializes in English literature and runs the school's drama club. Together, we make a supportive and nurturing family unit that values quality time, education, and open communication. Our family also enjoys traveling together, with recent trips including a ski vacation to Aspen and a cultural tour of Washington, D.C.
During my leisure time, I can often be found exploring the outdoors with my family, engaging in activities such as hiking in the picturesque Barton Creek Greenbelt, camping at the nearby Pedernales Falls State Park, and fishing on Lake Travis. As an avid reader, I enjoy immersing myself in the world of science fiction and fantasy, with a particular fondness for the works of Neil Gaiman and Ursula K. Le Guin. Additionally, I take pleasure in experimenting with gourmet cooking, exploring diverse cuisines, and sharing my culinary creations with my loved ones during our weekly family dinners.
In my personal and professional relationships, I appreciate sincerity, hard work, and dedication, qualities I strive to instill in my children and uphold in all aspects of my life.''

\textbf{Emily}
``Hi! I am Emily Johnson, and I am a 38-year-old accomplished architect. As a graduate of the Rhode Island School of Design, I have made a name for myself by designing sustainable buildings for prestigious clients. With over a decade of experience, I have become an indispensable asset to the award-winning firm, GreenScape Architects, where I have worked for the past six years. I am particularly fond of neoclassical and gothic architecture. I live in Portland, Oregon.
Residing in a charming, renovated Victorian house in a vibrant neighborhood, my home features four spacious bedrooms, intricately detailed walnut wooden staircases, and original black stained glass windows. The house is surrounded by a lush tomato garden and an outdoor seating area. My living space is a testament to my eye for African interior design, with a blend of modern minimalism and vintage charm.
I am a loving mother to my 7-year-old daughter, Sophie, whom I share with my ex-husband, James. Despite our differences, James and I maintain a healthy co-parenting relationship, ensuring Sophie grows up in a nurturing environment. My parents, Mary and Richard, live nearby and often lend a helping hand with childcare.
In my free time, I have a passion for photography, capturing the world around me through my unique perspective. My favorite photographer is Annie Leibovitz, whose work inspires my own photographic interests. I also enjoy practicing yoga, finding it to be a grounding and rejuvenating activity that helps me maintain a sense of balance amidst my busy life.
I am a fan of world cinema, with my all-time favorite movie being the independent film "Eternal Sunshine of the Spotless Mind." I appreciate the diverse storytelling techniques.
I have fond memories of my trip to Bangladesh, where I loved the vibrant culture and warm hospitality of the locals. I went for three months, from June to August of 1998. I visited the capital city of Dhaka and marveled at the architectural wonder of the Jatiya Sangsad Bhaban, the National Parliament House designed by Louis Kahn. I also ventured to the Sundarbans, the world's largest mangrove forest, where I was amazed by the rich biodiversity and had the opportunity to spot the elusive Bengal tiger from a safe distance. I cherished my time spent in the country, learning about its history, culture, and people.''

\textbf{Benjamin}
``Hey there! My name is Benjamin Martinez, and I am a 35-year-old environmental scientist living in the city of San Diego, California. Holding a Master's degree in Environmental Science from the University of California, Berkeley, I am passionate about preserving the planet for future generations. For the last eight years, I have been working at the non-profit organization, EarthGuard, where I lead research projects on ocean acidification and coral reef preservation.
To commute to work, I opt for an eco-friendly, multi-modal transportation route. I begin my journey by cycling from my home in Point Loma along a bike path, enjoying the ocean views as I pedal toward the Old Town Transit Center. Upon arriving, I secure my bicycle aboard bus number 36, which transports me to the Santa Fe Depot. From there, I board a commuter train that takes me to the EarthGuard office located near the Sorrento Valley station.
I reside in an eco-friendly home in the tranquil neighborhood of Point Loma. My residence is adorned with solar panels, energy-efficient appliances, and a vegetable garden that includes tomatoes, kale, and bell peppers, showcasing my commitment to reducing my environmental footprint.
I am married to my college sweetheart, Olivia, a talented graphic designer specializing in sustainable packaging. Together, we have a 4-year-old son, Lucas, who shares our love for nature and enjoys exploring the outdoors. We also have a Labrador retriever named Luna.
As an outdoor enthusiast, I enjoy hiking, mountain biking, and surfing, taking full advantage of Southern California's diverse natural landscapes, from the rolling hills of Balboa Park to the pristine beaches of La Jolla.
I am also an ardent music lover, with a diverse taste that ranges from classical compositions by Beethoven to indie rock bands like The National. I play the guitar and the piano and perform at local open mic nights hosted by Lestat's Coffee House. Attending music festivals, such as the annual San Diego IndieFest and Coachella, is one of my favorite music experiences.
In both my personal and professional life, I value integrity, empathy, and dedication. I am committed to making a positive impact on the world, ensuring that future generations continue to cherish and protect the planet.''

\textbf{Sarah}
``Hello! I am called Sarah Lee, and I am a 36-year-old graphic designer living in the vibrant city of Boston, USA. I moved to Boston a few years ago after receiving a job offer from a renowned advertising agency, where I eventually helped found the agency's design department. I have a passion for expressing my creativity through various forms of art. As a talented painter, I prefer using acrylic paints to bring my imaginative ideas to life on canvas. I draw inspiration from nature and often spend weekends exploring with my beloved Siberian Husky, Luna.
In addition to painting, I am an excellent cook and love experimenting with different cuisines. Some of my favorite recipes include homemade spinach and ricotta stuffed cannelloni, Thai green curry with shrimp, and a delectable Argentinean flan for dessert. I often turn to my extensive collection of cookbooks, such as "The Flavor Bible" by Karen Page and Andrew Dornenburg, online blogs like "Smitten Kitchen," and cooking shows, including "MasterChef," for inspiration and enjoy sharing my culinary creations with friends and family during dinner parties.
When I'm not in the kitchen or my art studio, I can be found at my local gym, "Pump Iron," where I am an enthusiastic fitness aficionado. I follow a disciplined workout routine, focusing on a mix of cardio exercises, strength training, and yoga. I usually go to the gym at 7:00 am and spend about an hour and a half there, ensuring I get a well-rounded workout. I am also a member of a nearby CrossFit center.
As a fan of strategy and critical thinking, I have amassed an impressive collection of board games, with my top three favorites being Settlers of Catan, Ticket to Ride, and Pandemic. I often organize game nights with my close friends, where we engage in friendly competition and enjoy each other's company.
My love for sports is apparent in my unwavering support for my favorite soccer teams, Everton and Wrexham. I never miss a match and can often be found at local sports bars or at home, cheering on my team with friends and fellow fans.''

\section{Scripted Questions}
\label{appendixB_scriptedqn}
The following shows the scripted general and specific questions for each fictional person. \\ \\

\noindent \textbf{Question Set 1 (William)} \\
\textit{General}: (1) ``I want to visit his family. Describe his family such as the names and ages.'' (2) ``We should hang out with this guy more. What are his hobbies? Where does he do his activities?''
\textit{Specific}: (1) ``I want to gift him a book for his birthday. Who are his favorite authors?'' (2) ``He is an inspirational father. What qualities does he teach his children?'' (3) ``I’d like to download his apps. What are the names of the apps he made?'' (4) ``His birthday is coming soon, let's surprise him. What’s his age and where does he live?''

\noindent \textbf{Question Set 2 (Emily)}\\
\textit{General}: (1) ``I want to get a house like her. Can you describe the house she has? Include as much detail.'' (2) ``What did she do on her recent trip? Describe it. I’d like to visit and do the same itinerary'' \textit{Specific}: (1) ``You heard about her daughter. What’s her daughter's name and age?'' (2) ``We should take her to a movie. What’s her favorite one?'' (3) ``She is a talented architect. What type of architecture does she like?'' (4) ``She told me many times but I forget. Who is her favorite photographer?''

\noindent \textbf{Question Set 3 (Benjamin)}\\
\textit{General}: (1) ``My friend is going to be working near him. What is his route to work? What are the things he sees on the way?'' (2) ``It seems like he likes music a lot. What instruments? Where does he perform and what festivals he goes to? Which artists does he like?'' \textit{Specific}: (1) ``Which university did he go to? I’d like to apply to it as well.'' (2) ``Why, should we buy vegetables when he grows them? What does he grow in his vegetable garden? (3) ``What is his wife's occupation? She is an ambitious woman.'' (4) ``My friend has been looking for a job. What's the name of the non-profit organization where he is working?''

\noindent \textbf{Question Set 4 (Sarah)}\\
\textit{General}: (1) ``She's an excellent cook! What are some of her favorite recipes, and where does she find inspiration for her cooking?'' (2) ``She seems to be a fitness enthusiast. What is her workout routine, which gym and what are her timings?''         
\textit{Specific}: (1) ``I've heard she's great at painting. What is her preferred medium?'' (2) ``She has an impressive collection of board games. What are her top three favorite board games?'' (3) ``What is her pet's name and what kind of pet does she have?'' (4) ``She's a fan of soccer and I wanted to buy her jerseys. Which teams does she support?''

\section{Example Queries}
\label{appendixC_examples}
\subsection{Baseline}
\textbf{Query}:  What were Sarah's favorite soccer teams? \\\textbf{Answer}:  Sarah's favorite soccer teams were Everton and Wrexham.\\
\textbf{Query}:   How does Benjamin get to work from the Santa Fe bus depot? \\
\textbf{Answer}:  Benjamin boards a commuter train from the Santa Fe Depot that takes him to the EarthGuard office located near the Sorrento Valley station.\\
\textbf{Query}:   What does Emily's house look like? \\ \textbf{Answer}:   Emily's house is a charming, renovated Victorian house in a vibrant neighborhood. It features four spacious bedrooms, intricately detailed walnut wooden staircases, and original and black stained glass windows. The house is surrounded by a lush tomato garden and an outdoor.\\
\textbf{Query}:    What vegetable starts does, Benjamin grow in his garden. \\
\textbf{Answer}:    Benjamin grows tomatoes, kale, and bell peppers in his garden.\\
\textbf{Query}:   What does William teaches kids? \\
\textbf{Answer}:  Benjamin boards a commuter train from the Santa Fe Depot that takes him to the EarthGuard office located near the Sorrento Valley station.\\

\subsection{Query Mode}
\textbf{Context:} went to Bangladesh,  Pictures of Bengal tiger, \\ \textbf{Query:}What else did she do? \\
\textbf{Answer}: Dhaka, Jatiya Sangsad Bhaban.\\
\textbf{Context:} pet's name is Luna. It's some sort of dog  \textbf{Query:}What type of dog does she have? \\
\textbf{Answer}: Husky.\\
\textbf{Context:} cycling going to get some bus 36. \\
\textbf{Query:} How does he go to work after Bus 36? \\ \textbf{Answer}:  Train.\\
\textbf{Context:} She's 7 years old I don't remember her name \\ \textbf{Query:} What was her name?\\ \textbf{Answer}: Sophie.\\
\textbf{Context:} San Diego Music Festival \\
\textbf{Query}: What is the place called again, at the bar? \\\textbf{Answer}: Lestat's.\\

\subsection{Queryless Mode}
\textbf{Context:} Hardware, sincerity and dedication, He made two apps. The names are \\
\textbf{Inferred Query:} What are the two apps that William Thompson created? \\ 
\textbf{Answer}: BuzzPal, FoodMingle\\
\textbf{Context:} now he's very fond of being outdoors. He likes hiking mountain biking and \\
\textbf{Inferred Query:}  What other outdoor activities does William Thompson enjoy? \\ 
\textbf{Answer}:  Camping, fishing.\\
\textbf{Context:}  okay, her house has full Venture bedroom and also and \\
\textbf{Inferred Query:}  What are the features of Emily Johnson's house, such as the walnut wooden staircases, blue stained glass windows, and outdoor seating area?  \\ 
\textbf{Answer}:  Walnut staircases, blue windows, outdoor seating.\\
\textbf{Context:} aw a scary tiger in the tiger is called, it's like a famous tiger called \\
\textbf{Inferred Query:}  What is the name of the famous tiger that was seen in the Sundarbans? \\
\textbf{Answer}: Bengal Tiger.\\
\textbf{Context:} She likes to paint on a canvas, And use. \\\textbf{Inferred Query:}  What type of paints does Sarah Lee prefer to use for her paintings? \\ 
\textbf{Answer}: Acrylic.\\

\section{Prompt Templates}
\label{appendixD_prompts}
\subsection{Contextual Query}
\begin{quote}
{\small \fontfamily{qcr}\selectfont
You are a helpful assistant that provides memory cues to a human. The human is engaged in a conversation with another human, and asks you in the middle for assistance.
The answer can be found in the relevant memories. If it is not found in the relevant memories, you should truthfully answer that you do not know the answer. \\
Relevant memories: \textit{<External Memories>} \\
The current context contains the conversation between the two humans. \\
Current context: \textit{<Current Context>} \\
The query is the question asked by the human to you. \\
Query: \textit{<Query>} \\
Answer: \textit{[Generated Answer]} 
}
\end{quote}
\subsection{Concise Suggestions}
\begin{quote}
{\small \fontfamily{qcr}\selectfont
Make the answer more concise, such that it only contains the words needed to answer the query. It should NOT contain any information that is already present in the current context. \\
Current context: \textit{<Current Context>} \\
Query: \textit{<Query>} \\
Answer: \textit{<Retrieved Answer>} \\
Concise answer: \textit{[Generated Answer]} \\
}
\end{quote}
\subsection{Queryless Search}
\begin{quote}
{\small \fontfamily{qcr}\selectfont

You are an assistant interface between user and a memory system. The user is engaged in a conversation with another human, and asks you in the middle for assistance.
The assistant frames a query that the user would like to ask the memory system next at the end of the conversation.
The recent conversation between the two humans is related to the relevant memories. The answer that the user would like to retrieve would not be in the recent conversation. 
The query should be very relevant to the end of the last sentence of the recent conversation. \\
Recent conversation: \textit{<Current Context>} \\
What do you think that the user would like to ask the memory system to finish or clarify his last sentence? \\
Query: [Generated Query]
}
\end{quote}

\section{Questionnaires}
\label{appendixE_questionnaires}
\subsection{User experiences and Perception}
We measured eight aspects using a 7-point Likert scale (1=strongly disagree, 7=strongly agree). 

\begin{enumerate}
    \item \textbf{Length of Responses}: ``I felt that the length of the answers was appropriate.''
    \item \textbf{Adaptiveness of the System}: ``I felt that the system adapted to my needs in the conversation.''
    \item \textbf{Interruption to Conversation}: ``The device manipulation by me interrupted the conversation.''
    \item \textbf{Helpfulness of Response}: ``The answers from the system were helpful.''
    \item \textbf{Usefulness}: ``The system would be useful in my everyday life.''
    \item \textbf{Politeness}: ``I felt it was polite to use the system during the conversation.''
    \item \textbf{Naturalness}: ``I acted naturally at all times while focusing on the researcher’s face and using the system.''
    \item \textbf{Ease of Ignoring the Device}: ``It was easy to ignore the fact that I was wearing the device.''
\end{enumerate}

\subsection{Conversation Quality}
We measured six aspects using a 7-point Likert scale (1=strongly disagree, 7=strongly agree). ~\cite{cai2023paraglassmenu}: 
\begin{enumerate}
    \item \textbf{Listening to the Conversational Partner}: ``When the other person was speaking, I was always listening to them.''
    \item \textbf{Concentration on the Conversation}: ``I was always concentrating on the conversation.''
    \item \textbf{Attention Towards Conversation Partner}: ``When I was speaking, my attention was towards the other person.''
    \item \textbf{Eye Contact}: ``When I was speaking, I maintained eye contact.''
    \item \textbf{Naturalness}: ``I acted naturally at all times during the conversation.''
    \item \textbf{Feeling Relaxed}: ``I felt relaxed during the conversation.''
\end{enumerate}
\subsection{Task Performance/Recall Ability}
We measured three aspects using a 100-point slider scale. It involved: \begin{enumerate}
    \item \textbf{Confidence in Memory}: ``I was confident in my ability to recall the information of the person while answering the questions.''
    \item \textbf{Difficulty in Recall}: ``I found it difficult in recalling the information of the person.''
    \item \textbf{Recalled Relevance}: ``I recalled all the relevant information of the person with respect to the question.''
\end{enumerate}

\appendix

\end{document}